\documentclass[acmsmall,screen,nonacm]{acmart}
\usepackage{amsfonts}
\usepackage{textcomp}
\usepackage[ruled,vlined,linesnumbered]{algorithm2e}
\usepackage{pifont}
\usepackage{listings}
\usepackage{multirow}

\AtBeginDocument{%
  }

\SetKw{Instantiate}{Instantiate}
\SetKw{Connect}{Connect}

\begin{document}

\title{A Novel FPGA Delay Structure to Improve Ease of Implementation and PUF Metrics for FPGA-Based Arbiter PUFs}

\author{Syed Talha Imam}
% \email{simam2025@fau.edu}
\affiliation{%
  \institution{Department of Electrical Engineering and Computer Science, Florida Atlantic University}
  \city{Boca Raton}
  \state{Florida}
  \country{USA}
}

\author{Ayesha Javaid}
% \email{ayesha.javaid@lums.edu.pk}
\affiliation{%
  \institution{Independent Researcher}
  % \city{Boca Raton}
  \country{USA}
}

\author{Arslan Munir}
\correspondingauthor
% \email{arslanm@fau.edu}
% \orcid{0000-0002-3126-8945}
\affiliation{%
  \institution{Department of Electrical Engineering and Computer Science, Florida Atlantic University}
  \city{Boca Raton}
  \state{Florida}
  \country{USA}
}

\begin{abstract}
Arbiter PUFs are delay-based Physical Unclonable Functions (PUFs) that are one of the foundational Strong PUFs in reconfigurable hardware; however, the difficulty of managing routing constraints precisely on Field-Programmable Gate Array (FPGA) toolchains to make the response sensitive to random wire delay variations make them difficult to implement practically with acceptable PUF metrics.
In this paper, we propose a novel delay structure built entirely using FPGA Configurable Logic Block (CLB) primitives including Look-Up Tables (LUTs) and carry-chains with the aim of making the design easy to implement with software-automated tuning while providing acceptable PUF metrics. 
We utilize these delay structure blocks to create tunable delay lines, with a software interface to tune the delay lines until acceptable PUF metrics have been achieved.
We then combine these delay lines to create a novel arbiter PUF architecture, called Helix Arbiter PUF (HAPUF), that provides us near-ideal PUF metrics in uniqueness, reliability, and uniformity with a single software tuning run. 
We also show a variant of our delay block where each delay line has a unique seed-permutation associated with it.
The response is unique to the seed-pair used in case of different seeds for each delay line and if the same seeds are used for both delay lines, the behavior is similar to the standard implementation along with possible challenge-obfuscation. 
We finally present our tuning software algorithm, which provides us tuning to implement the particular PUF with suitable metrics on each device.
We test our design on multiple Xilinx Arty A7-100t FPGA boards and also test metrics under voltage and temperature variation.
Finally, we show that our design provides acceptable metrics with a single implementation flow iteration with software-based tuning compared to other delay blocks in literature.

\end{abstract}

% \begin{CCSXML}
% <ccs2012>
%    <concept>
%        <concept_id>10010583.10010682.10010712.10010713</concept_id>
%        <concept_desc>Hardware~Best practices for EDA</concept_desc>
%        <concept_significance>500</concept_significance>
%        </concept>
%    <concept>
%        <concept_id>10002978.10003001.10003599</concept_id>
%        <concept_desc>Security and privacy~Hardware security implementation</concept_desc>
%        <concept_significance>500</concept_significance>
%        </concept>
%    <concept>
%        <concept_id>10010520.10010521.10010542.10010543</concept_id>
%        <concept_desc>Computer systems organization~Reconfigurable computing</concept_desc>
%        <concept_significance>500</concept_significance>
%        </concept>
%  </ccs2012>
% \end{CCSXML}

% \ccsdesc[500]{Hardware~Best practices for EDA}
% \ccsdesc[500]{Security and privacy~Hardware security implementation}
% \ccsdesc[500]{Computer systems organization~Reconfigurable computing}

\keywords{Delay Line, Field-Programmable Gate Array, Hardware Security, Internet of Things, Tuning, Physical Unclonable Functions,  Resource-Constrained Devices, Strong PUFs, Arbiter PUFs}

\maketitle

%%%%%%%%%%%%%%%%%%%%%%%%%%%%%%%%%%%%%%
%%%%%%%%%%%% SECTION I %%%%%%%%%%%%%%%
%%%%%%%%%%%%%%%%%%%%%%%%%%%%%%%%%%%%%% 
\section{Introduction}
\label{sec:introduction}
Physical Unclonable Functions (PUFs) are a promising technology that are naturally occurring fingerprints in hardware circuits.
They attempt to provide an alternative to the use of Public-Key Cryptography (PKC) for authentication and key-exchange~\cite{suh2007physical}. 
PUFs are electronic circuits such as the signal delay of a wire \cite{gassend2002silicon} or the startup value of an SRAM cell \cite{holcomb2009power} that are designed with the exact same blueprint in multiple productions of the same device.
However, due to the inter-process variation at time of manufacturing of semiconductor devices and integrated circuits at the foundry, each device has a unique value associated with their instance of the PUF.
This serves as the fingerprint for each manufactured instance of the device. 
The authenticator then may enroll a response of the PUF and then query it later to authenticate the device. 
Alternatively, the PUF value may be used as a secret key as its response is random to each device.
These secret keys can then be combined or utilized to obtain a symmetric secret key for symmetric encryption.

By providing an alternative to traditional public-key authentication, PUFs provide an effective way to perform lightweight authentication of resource-constrained Internet-of-Things (IoT) devices without the need for storing secrets in the Non-Volatile Memory (NVM). 
PUFs must also have a suitable number of Challenge-Response Pairs (CRPs) so that an authenticator may be able to authenticate the PUF as many times as they wish through the lifetime of the system.
In this aspect, PUFs are categorized into two categories \cite{ruhrmair2010modeling}:
\begin{itemize}
    \item Strong PUFs: PUFs that have a large challenge space. 
    Examples are delay-based PUFs such as arbiter PUFs whose challenge space grows exponentially with the number of delay stages. 
    They are mainly used for IoT device authentication through challenge–response verification.
    \item Weak PUFs: PUFs that have a comparitively smaller challenge space such as memory-based PUFs.
    They are mainly used for cryptographic key generation for secure boot and data encryption.
\end{itemize}
One strong PUF that we will be focusing on this paper is the Arbiter PUF (APUF) \cite{gassend2002silicon}. 
The APUF consists of two delay lines of muxes that swap paths based on the challenge bit to each mux stage. 
At the final stage, there is an arbiter which determines which delay line is faster.
The challenge provided to the APUF are the select lines of the muxes in each stage of the delay line.
The more stages in the delay lines, the more bits in the challenge.
The output of the the final-stage arbiter is the response of the PUF.
The response is a single bit response from the arbiter, but a multiple bit response can be obtained by combining the responses of multiple challenges.

Field-Programmable Gate Arrays (FPGAs) are popular devices in both the IoT sphere and compute-intensive areas \cite{FPGAXilinx}.
They are reconfigurable hardware which allows the user to program digital circuits onto the device \cite{reconfigurableestrin}.
Although their performance is not at the level of foundry manufactured hard-silicon, they are still popular for their fast time to market, and for the deployment of accelerated hardware. 
FPGAs have also been utilized to implement PUFs, especially the APUF.
However, APUF implementations on FPGAs suffer from several shortcomings.
These include difficulty in implementation, poor reliability,  and poor uniqueness due to asymmetric routing on the FPGA routing matrix.

We propose a novel delay line design using FPGA primitives that addresses these challenges.
Using these delay lines, we present a novel FPGA APUF design that we name Helix-Arbiter-PUF (HAPUF).
Our design is motivated by ease of implementation along with near-ideal metrics in uniformity, uniqueness and reliability. 
The main contributions of this paper through this design are:
\begin{itemize}
    \item {A novel FPGA delay-block structure using carry-chains and Lookup-Tables (LUTs).}
    \item {An improved FPGA APUF design with delay lines using this delay block structure.}
    \item {A seed-based delay line switching configuration to create unique instances based on the uniquely different seed-pairs for the two delay lines.}
    \item {A software tuning alogorithm to provide PUF instances with acceptable PUF metrics with a single software tuning-run.}
    \item {Thorough reliability testing of the PUF design for voltage and temperature variation and comparison to other proposed delay blocks.}
\end{itemize}

The remainder of this paper is organized as follows. Section~\ref{sec:backgnd&rw} discusses the background and related work for this paper. 
Section~\ref{sec:proposed} presents the novel delay block design, software tuning and HAPUF design. 
Section~\ref{sec:methodology} goes over the implementation and experimentation methodology for this design. 
Finally, Section~\ref{sec:Conclusion} provides a summary for the conclusion of the paper.

%%%%%%%%%%%%%%%%%%%%%%%%%%%%%%%%%%%%%%%%%
%%%%%%%%%%% FIGURE 1 %%%%%%%%%%%%%%%%%%%%
%%%%%%%%%%%%%%%%%%%%%%%%%%%%%%%%%%%%%%%%%
\begin{figure}[t]
\centering
\includegraphics[width=\columnwidth]{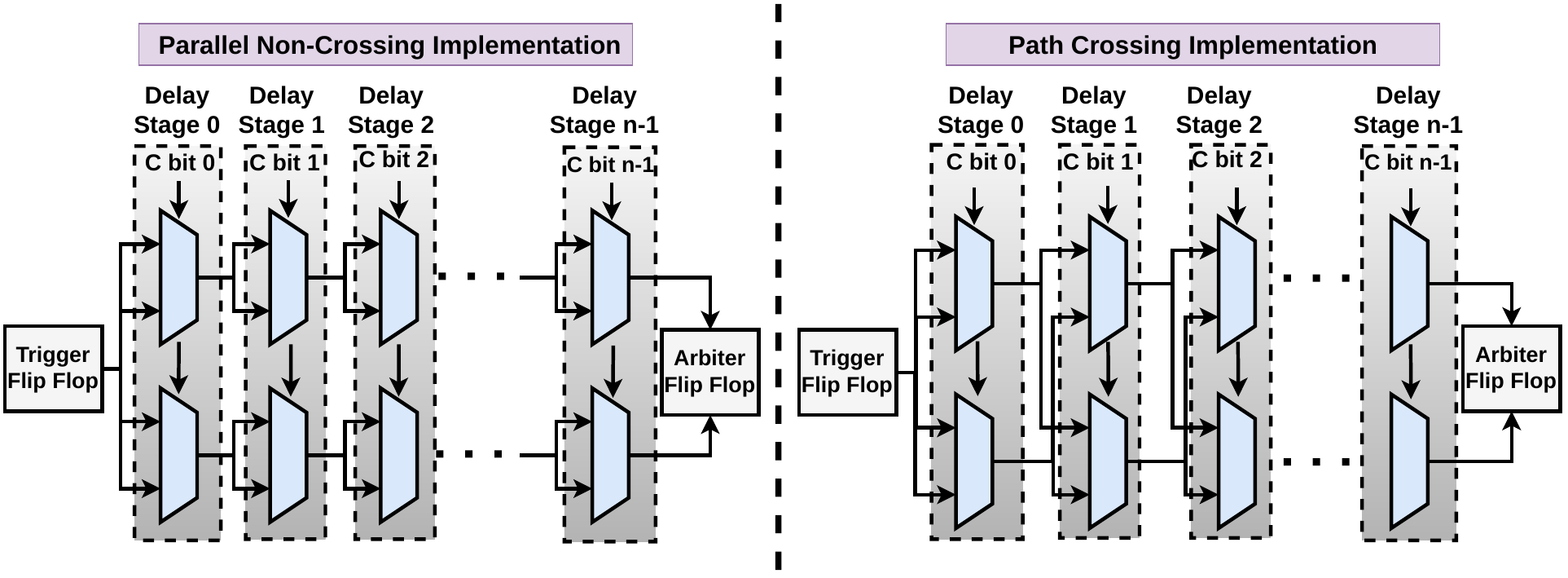}
\caption{Standard Arbiter PUF design. C is challenge, and R is response.}
\Description{A figure showing the layout of a standard Arbiter PUF designs with the trigger sending the signal through the two delay lines. It shows two different variants. One includes challenge based delay line crossing by the two signals while in the other the signal remains in its delay line.}
\label{fig:std_arbpuf}
\end{figure}
%%%%%%%%%%%%%%%%%%%%%%%%%%%%%%%%%%%%%%%%%
%%%%%%%%%%%%%%%%%%%%%%%%%%%%%%%%%%%%%%%%%
%%%%%%%%%%%%%%%%%%%%%%%%%%%%%%%%%%%%%%%%%

\section{Background and Related Work}
\label{sec:backgnd&rw}
First, we introduce some mathematical notation that will be used throughout this paper.
Vector quantities are denoted by an arrow. For example, $\vec{c} = [c_1,c_2,\ldots,c_n]$ denotes an n-bit challenge vector, where $c_i \in {0,1}$ is the i-th challenge bit.
The dot operator $\cdot$ denotes the inner product between two vectors. 
For vectors $\vec{x},\vec{y}\in\mathbb{R}^n$, the dot product is defined as $\vec{x}\cdot\vec{y}=\sum_{i=1}^{n}x_i y_i$.

\subsection{Background}
\label{subsection:background}
\subsubsection{Arbiter PUFs on FPGAs:}
\begin{figure}[b]
\centering
\includegraphics[width=\columnwidth]{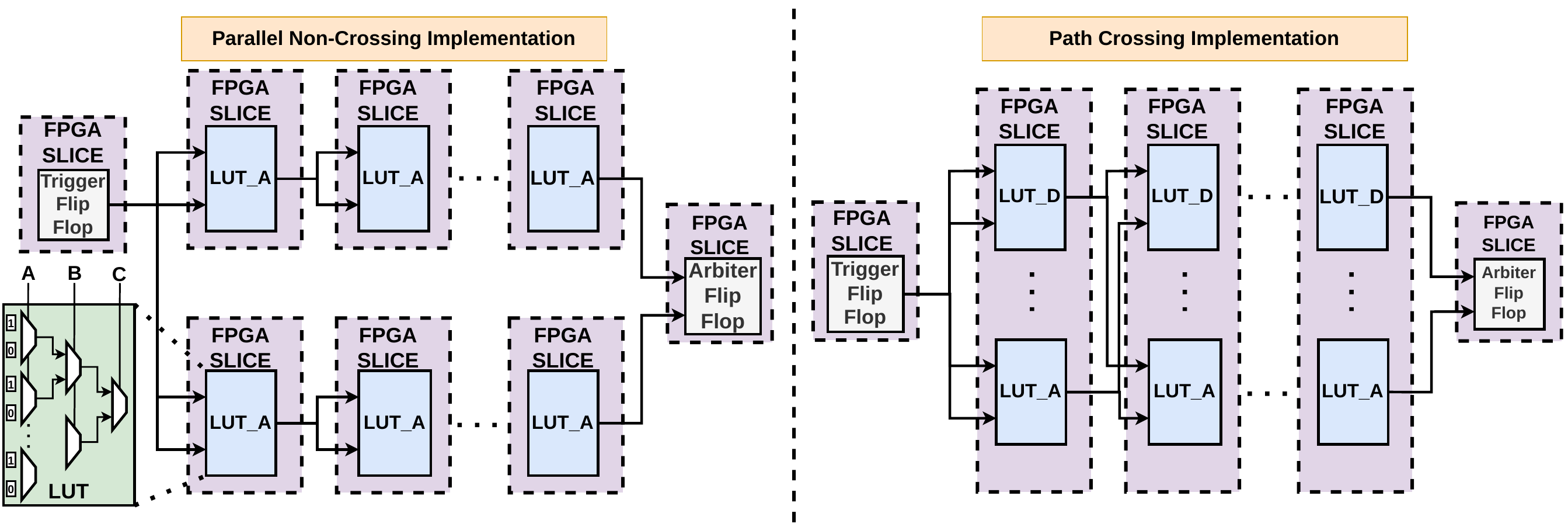}
\caption{Standard Arbiter PUF design implementation on a typical FPGA. LUT is Lookup Table.}
\Description{Two two standard arbiter PUF implementation of the line crossing and non-crossing variants are shown how they map to FPGA hardware. Figure shows how multiplexers are mappes to LUTs in slices and trigger and arbiter are mapped to flip flops.}
\label{fig:std_fpga_arbpuf}
\end{figure}

As discussed, APUFs are strong PUFs that compare the signal delay of two competing paths and output the result of the race.
The challenge switches the path in the delay-stages of the two paths, and response is the result for the two specific paths against those challenges.
For implementing APUFs on Application Specific Integrated Ciruits (ASICs), the designer has the liberty of placing the two delay lines symmetrically to ensure there is no bias and the output is determined by random inter-process variations at the time of manufacturing \cite{majzoobi2010fpga}.
For an FPGA implementation, the APUF must be mapped onto the primitive reconfigurable hardware blocks available on the FPGA and utilize the routing matrix provided in the FPGA matrix~\cite{majzoobi2010fpga}.
Figure~\ref{fig:std_arbpuf} shows two standard implementation of APUFs: the parallel non-crossing implementation and the path-crossing implementation.
We will differentiate between them in detail later.
To implement these designs onto an FPGA, it must be implemented on the Lookup Tables (LUTs) and Slices on the FPGA matrix. 
Figure ~\ref{fig:std_fpga_arbpuf} shows the FPGA implementation of the two APUFs.

%%%%%%%%%%%%%%%%%%%%%%%%%%%%%%%%%%%%%%%%%
%%%%%%%%%%% FIGURE 2 %%%%%%%%%%%%%%%%%%%%
%%%%%%%%%%%%%%%%%%%%%%%%%%%%%%%%%%%%%%%%%
%%%%%%%%%%%%%%%%%%%%%%%%%%%%%%%%%%%%%%%%%
%%%%%%%%%%%%%%%%%%%%%%%%%%%%%%%%%%%%%%%%%
%%%%%%%%%%%%%%%%%%%%%%%%%%%%%%%%%%%%%%%%%

\subsubsection{PUF Metrics:}
There are multiple metrics that are of interest when determining the quality of a PUF \cite{hori2010quantitative}.
A PUF must be reliable in its response, unique between multiple instances, and have a uniformly distributed response.
A list of metrics that we will be focusing on and measuring is detailed next:

\paragraph{Uniqueness (Inter-Chip Hamming Distance):} Uniquess compares how responses against multiple chip instances of the same PUF compare. It is calculated using the following equation:
\begin{equation}
    Uniqueness = \frac{2}{k(k-1)} \sum\limits_{u=1}^{k-1} \sum\limits_{v=u+1}^{k} \frac{HD(R_u, R_v)}{n} \times 100\%
    \label{eq:puf_uniqueness}
\end{equation}
Here $k$ is the total number of chips evaluated, $n$ is the number of bits in the response, and $R_u$ and $R_v$ are the responses of PUF chips $u$ and $v$ being considered.
$HD$ is the hamming distance of the two responses.
Ideal uniqueness is 50\% which indicates that the PUF is completely unique when compared to all other physical instances of that PUF.

\paragraph{Reliability (Intra-Chip Hamming Distance):} Reliability measures how much a PUF instance's response deviates from the baseline across multiple measurements.
These measurements may be made under varying conditions such as varying temperature and voltage.

\begin{equation}
    Reliability = 100\% - (\frac{1}{m} \sum\limits_{i=1}^{m}\frac{HD(R, R_i)}{n} \times 100\%)
    \label{eq:puf_reliability}
\end{equation}
Here $m$ is the number of total measurements performed.
$R$ is the golden-reference baseline response, and $R_i$ is the response of the current measurement.
An ideal reliability of 100\% indicates that the response is always identical for a particular PUF chip.

\paragraph{Uniformity :} Uniformity measures the proportion of 1s in the response of a single PUF instance.
\begin{equation}
    Uniformity = \frac{1}{n}\sum\limits_{i=1}^{n}R_{i} \times100\%
    \label{eq:puf_uniformity}
\end{equation}
An ideal uniformity of 50\% is aimed for so that the PUF has a response with an equal proportion of 1s and 0s.

\subsubsection{Delay Model for FPGA Implementation:}
In an ASIC realization, the APUF involves path-swapping between the two delay paths.
Most FPGA implementations of the APUF fix the two paths to the same delay lines (e.g.\ a single X or Y coordinate column) for the full stage chain~\cite{majzoobi2014, mahalat2021PCS, sahoo2015towardsidealapuf}, rather than swapping the delay lines at each stage.
This is done since path-swapping to a different delay line in the FPGA matrix can introduce routing bias that can dominate the design. 
\autoref{fig:std_arbpuf} shows both implementations. 
We will discuss the delay models for both.
\paragraph{Parallel Non-Crossing Implementation:}
As~\autoref{fig:std_arbpuf} and~\autoref{fig:std_fpga_arbpuf} shows, in the parallel non-crossing implementation, each stage involves a constant wire-delay connecting two stages, and the top or bottom path of the mux chosen with the challenge bit for that stage.
Let $P_0, P_1$ denote these two fixed tracks.
At delay line $j$ and stage $i$, each track accumulates one of two possible delay contributions, $B_{j,i}$ and $T_{j,i}$, selected by the challenge bit $c_i$ and constant inter-stage wiring delay $K_{j,i}$ common to both selections.
Representing the delay contributed at stage $i$ at line $j$ as $\lambda_{j, i}$:
\begin{equation}
    \lambda_{j,i} = T_{j,i}(1-c_i)+B_{j,i}c_i+K_{j,i}
    \label{eq:stagedelay}
\end{equation}
Apart from the stage delays the two lines have constant delays from the trigger flop to the first-stage and to the arbiter flop from the last-stage.
We combine these constant delays in each line and represent them as $d_j$.
Now creating an expression for the delay of the lines $P_0$, and $P_1$:
\begin{equation}
    P_j = \sum_{i=1}^n \lambda_{j, i} + d_j
    \label{eq:delaypath0}
\end{equation}
Expanding $\lambda_{j,i}$:
\begin{equation}
    P_0 = \sum_{i=1}^n\Big[T_{0,i}(1-c_i)+B_{0,i}c_i+K_{0,i}\Big] + d_0
    \label{eq:delaypath0}
\end{equation}

\begin{equation}
    P_1 = \sum_{i=1}^n\Big[T_{1,i}(1-c_i)+B_{1,i}c_i+K_{1,i}\Big] + d_1
    \label{eq:delaypath1}
\end{equation}
These delay equations may be separated by the challenge-affected delay and constant delay:
\begin{equation}
    P_j = \sum_{i=1}^n\Big[T_{j,i}(1-c_i)+B_{j,i}c_i\Big] + \Big\{\sum_{i=1}^{n}K_{j,i} + d_j \Big\}
    \label{eq:delaypathx}
\end{equation}
The difference of the two delay paths $\Delta$ is then calculated as:
\begin{equation}
    \Delta = P_1 - P_0
    \label{eq:deltasimple}
\end{equation}
Separating, the challenge-based delays and constant delay in this subtraction into variables $\delta$ and $\kappa$:
\begin{equation}
    \delta = \sum_{i=1}^n\Big[(T_{1,i} - T_{0,i})(1-c_i) + (B_{1,i}-B_{0,i})(c_i)\Big]
    \label{eq:delaychallenge}
\end{equation}

\begin{equation}
    \delta = \sum_{i=1}^n\Big[T_{1,i} - T_{0,i}\Big] + \sum_{i=1}^nc_i\Big[(T_{0,i} - T_{1,i} + B_{1,i} - B_{0,i})\Big]
    \label{delta_expand}
\end{equation}
We can remove the constant sum from the $\delta$ equation and represent as a separate variable $\kappa_{T}$:
\begin{equation}
    \kappa_T = \sum_{i=1}^n\Big[T_{1,i} - T_{0,i}\Big]
    \label{eq:kapptT}
\end{equation}
The constant term of the difference $\Delta$ is now: 

\begin{equation}
    \kappa = \kappa_T + \sum_{i=1}^n\Big[K_{1,i} - K_{0,i}\Big] + (d_1 - d_0)
    \label{eq:delayconstant}
\end{equation}
We can reframe the $\Delta$ equation as now:

\begin{equation}
    \Delta = \sum_{i=1}^nc_i\Big[(T_{0,i} - T_{1,i} + B_{1,i} - B_{0,i})\Big] + \kappa
    \label{eq:delta_bias}
\end{equation}
With the introduction of a weight variable $w$:
\begin{equation}
    w_i = \Big[(T_{0,i} - T_{1,i} + B_{1,i} - B_{0,i})\Big]
    \label{eq:weight}
\end{equation}
The $\Delta$ equation becomes:

\begin{equation}
    \Delta = \kappa + \sum_{i=1}^nc_iw_i = \kappa + \vec{c} \cdot \vec{w}
    \label{eq:delta_weight}
\end{equation}
The response is then:
\begin{equation}
    R= 
    \begin{cases}
        1\text{ if }\Delta > 0 \\
        0\text{ if }\Delta \leq 0
    \end{cases}
    \label{eq:stdapuf_resp}
\end{equation}
From the above model key observations can be made about the non-crossing FPGA APUF implementation in literature:
\begin{itemize}
    \item The standard FPGA APUF delay model is affine in the raw challenge bits $\vec c$, as shown in~\eqref{eq:delta_weight}: linear in each $c_i$ with a constant offset $\kappa$. 
    This affine model is supported by the works of Lim et al.~\cite{lim2005_affine} and Majzoobi et al.~\cite{majzoobi2010fpga}.
    \item The constant term $\kappa$, which we define as bias, shifts $\Delta$ uniformly across all $2^n$ challenges independent of $\vec c$. 
    A nonzero $\kappa$ of either sign therefore deviates the response distribution away from the ideal $50\%$ uniformity~\eqref{eq:puf_uniformity}.
    To implement this PUF on FPGAs, this bias must be eliminated which is a difficult task with FPGA routing limitations.
    \item A single bit flip in $c_i$ changes $\Delta$ by exactly $w_i$, independent of the values of every other challenge bit. 
    The marginal effect of each bit is therefore fixed. 
    A stage with anomalously large $w_i$ can dominate a response. 
    Similarly, an anomalously small $w_i$ may not affect the response sufficiently to affect the response. Stefani et al.~\cite{stefani2024strong} focuses on single-bit flip as a metric among others.
\end{itemize}

The bias term in the final equation can be quite large as the derivation showed.
Since the routing between slices is much larger than the routing in FPGA LUT paths, this bias is a common culprit in making this implementation difficult to implement, and prone to poor metrics.
Our design will continue the good practice of this design in not crossing delay lines to avoid routing asymmetry but avoid its shortcomings with bias and fixed effects of each challenge bit.

\paragraph{Path-crossing Implementation:}

The other model of APUF implemented on FPGAs follows the path-crossing implementation as the ASIC implementation. \autoref{fig:std_arbpuf} and~\autoref{fig:std_fpga_arbpuf} shows the model and the FPGA implementation of this model. 
For this variant, we refer to the mathematical model derived by Lim et~al.~\cite{lim2005_affine} for the ASIC realization of Gassend et~al.~\cite{gassend2002silicon}.
Like the non-swapping model, a common trigger flop launches the signal which then runs on two wire tracks in parallel.
Unlike the non-swapping model, at each of $n$ stages a switch selects between straight and crossed configurations. 
Let $\Delta_i$ denote the top-minus-bottom accumulated delay difference after stage $i$, and let $\alpha_i, \beta_i$ denote the per-stage intrinsic delay asymmetries under the straight and crossed configurations respectively. 
The recurrence is
\begin{equation}
  \Delta_i =
  \begin{cases}
    \Delta_{i-1} + \alpha_i, & c_i = 0 \\
    -\Delta_{i-1} + \beta_i, & c_i = 1
  \end{cases}
  \label{eq:apuf-recurrence}
\end{equation}
The sign inversion under $c_i = 1$ captures the physical path swap.
When the tracks exchange, top-path moves to the bottom and vice versa.
Therefore $\Delta_i$ which represents top-minus-bottom for stage $i$ changes sign.
$\alpha_{i}$ and $\beta_{i}$ are added to the delay difference depending on whether a crossing is performed or not as those are the intrinsic delays associated with each path.
Unrolling \eqref{eq:apuf-recurrence} using the sign selector $(1 - 2c_i) \in \{+1, -1\}$ and defining the parity feature vector $\vec \Phi = [\Phi_0, \Phi_2, ... , \Phi_n]$ for $n$ delay stages:
\begin{equation}
  \Phi_i(\vec c) = \prod_{k=i}^{n}(1 - 2c_k),
  \qquad \Phi_{n+1} = 1,
  \label{eq:phi-def}
\end{equation}
gives the well-known linear additive delay
model~\cite{lim2005_affine}:
\begin{equation}
  \Delta = \vec w \cdot \vec\Phi(\vec c),
  \qquad 
    R= 
    \begin{cases}
        1\text{ if }\Delta > 0 \\
        0\text{ if }\Delta \leq 0
    \end{cases}
  \label{eq:apuf-closedform}
\end{equation}
where the weights $\vec w = [w_1, w_2, ...,w_n]$ are linear combinations of the intrinsic stage delays $\alpha_i, \beta_i$~\cite{lim2005_affine} for each delay stage. 
This model is the standard for analysis for crossing implementations of arbiter PUF.

Analyzing ~\eqref{eq:apuf-closedform}, we can see two properties that distinguish it from the non-swapping FPGA model of~\eqref{eq:delta_weight}. 
First, $\Delta$ is a degree-$n$ polynomial in raw $\vec c$ but linear in the parity vector $\vec\Phi(\vec c)$. 
Second, flipping any single bit $c_j$ inverts the sign of every $\Phi_i$ for $i \le j$, propagating the change through $j$ terms of the sum simultaneously, whereas in the non-swapping FPGA model, a single-bit flip alters exactly one term of $\Delta$. 
Both properties are due to the sign inversion $-\Delta_{i-1}$ in~\eqref{eq:apuf-recurrence} which is caused by the physical wire crossing between the two delay-line tracks at each swapped stage.
This improves on the fixed contribution of a bit-flip by $w_{i}$ shortcoming as in the non-crossing implementation.
Further bias is also not accumulated in every stage due to path swapping.
However, due to the asymmetric routing when crossing delay lines in the FPGA matrix, the weight vector $\vec w$ in this case, is dominated by asymmetry so random variation in the delay difference of the two paths does not contribute as much to the response.
This causes poor uniqueness in most implementations.
Unlike the ASIC implementation, where every path going into the next stage, crossing or non-crossing, is designed to have an identical signal delay, the FPGA routing matrix will be non-uniform when signals are made to cross the delay lines.
Therefore, while the non-crossing implementations both the top and bottom delay lines have the same routing excluding bias, and the response is only dependent on the random variation, in the path-crossing implementation the response may depend on the different routes taken due to different routing from crossing rather than random variation.
Poor uniqueness is especially common if the path crossing implementation is implemented with a different SLICE column for each delay line.
This is because crossing SLICE columns introduces a much different routing path where routing asymmetry can dominate random variation.
Although this model gives good results in uniformity, uniqueness can be significantly more difficult to implement.

Our design will again take the good findings of this design in eliminating bias and in eliminating independent contribution by $c_{i}$.
But we will avoid delay-line-swapping to preserve uniqueness while also providing a similarly complex delay-model for desired uniformity.

%%%%%%%%%%%%%%%%%%%%%%%%%%%%%%%%%%%%%%%
%%%%%%%%%%% SECTION II %%%%%%%%%%%%%%%%
%%%%%%%%%%%%%%%%%%%%%%%%%%%%%%%%%%%%%%% 

\subsection{Related Work}
Recent works have attempted to improve upon the delay model of the FPGA APUF and enhance its design to improve PUF metrics and provide an architecture that is easier to implement. 
In our literature review, we will classify the related literature into two sections. 
One will go over related works that make architectural improvements to the delay model, and the other will go over literature that does algorithmic and obfuscation improvements to the PUF.

\subsubsection{Architectural improvements to the Delay Model:}
Several works have identified the linearity of the standard model and have aimed to improve it. Gassend et al.~\cite{gassend2004identification} introduced the feed-forward APUF (FF-APUF) that evaluates intermediary responses and feeds them forward as a challenge. 
However, since FF-APUF will be built using non-crossing or crossing implementations as discussed in~\autoref{subsection:background}, they will have the same implementation difficulties as their building-block arbiter PUFs. 
The XOR-APUF introduced in~\cite{suh2007physical} and the Double-Arbiter PUF~\cite{machida2015DAPUF} also use the same PUFs as their foundational blocks and so will run into similar problems. 
Mahalat et al.~\cite{mahalat2021PCS} implement a path-changing switch (PCS) using a slice of the FPGA.
The PCS makes improvements over the standard arbiter PUF PDL implementations. 
They use a 4x1 mux comprising of a complete slice to provide significantly different mux paths in delay, and they also avoid crossing between delay lines, however, as we will show in~\autoref{sec:methodology}, they also need careful routing for ideal uniqueness.
Hatti et al.~\cite{hatti2022dffxor} implements a double feed-forward arbiter PUF on an FPGA, but it also has the same problems as with standard FF-APUF and XOR-APUF. 
Gu et al.~\cite{gu2019ffapuf} implement an APUF on an FPGA whose delay line utilizes the slice flip-flops to provide better delay process-variations, however, their implementation is similar to the non-crossing PCS implementation with a 4-way switch rather than a 2-way, however the authors mention that implementation requires careful routing constraints and routing asymmetries in the 4-way switch shows in their uniqueness of $41.53\%$.
Aknesil et al.~\cite{aknesil2021_4x4apuf} implement an APUF on an FPGA where they implement a 4x4 switch instead of a 2x1 switch as in standard APUFs.
However, they are also unable to demonstrate ideal uniqueness showing how FPGA routing can override the process variations that make PUFs a unique fingerprint.
The authors of~\cite{yavuz2024_24dapuf} similarly also create a delay-stage switch with an increase in challenge inputs and and increased stage outputs.
This is a improvement over the baseline designs we discussed however, they utilize two slices per delay stage and require careful routing for metrics. 
Wang et al.~\cite{wang2024quantizeapuf} implement an FPGA APUF whose response is the quantized delay difference between the two paths rather than a single response. 
However, they permit a 3-bit tolerance in the response to achieve good reliability. 
Additionally, the design has the same difficulties in implementation as the standard APUF.
The authors of~\cite{xu2023obfuscate} create a delay stage switch whose connections to the next-stage are random and obfuscated to the adversary.
This design's security is tied to the secret that is the random connections between stages.
However, the design will have the same bias problems as the non-crossing implementation in~\autoref{subsection:background} and the authors do thoroughly careful routing which is difficult to implement.

\subsubsection{Algorithmic Enhancements:}
In a different manner, obfuscations and challenge or response processing has also been utilized to improve the performance of APUFs. 
Majzoobi et al.~\cite{majzoobi2008lightweight} and Gassend et al.~\cite{gassend2002silicon} introduced the hashing and processing of challenges and responses to obfuscate the output of PUFs  and break the linear input to the APUF.
The interpose-PUF proposed in~\cite{nguyen2019interpose} implemented two PUFs, with the response of the first PUF interposed into the challenge of the second PUF. 
This design uses more FPGA resources, and can have cascaded reliability problems due to the chained PUF setup. 
Zalivaka et al.~\cite{zalivaka2018lsfrbilbodnn} wrap their FPGA APUF with LFSR based challenge-inputs and a Deep Neural Network based unstable, metastable and stable response classifier. 
This incurs computational overhead and does not solve the inherent difficulty of implementing the APUF on FPGAs.
Huang et al.~\cite{huang2025pfopuf} implement two APUFs with the intermediate responses of the first being used to manipulate the challenge of the second PUF. 
Both PUFs intermediate responses are then used in an obfuscation module to switch operating modes. 
The authors of~\cite{ma2018weakstrong} use the response of multiple weak PUFs to manipulate the challenge being fed to the eventual strong PUF. Hou et al.~\cite{hou2019weakLFSRstrong} similarly use multiple weak PUF and an LFSR to provide the challenge for the eventual strong PUF. 
Both these implementations have multiple points of failure from a reliability perspective.

%%%%%%%%%%%%%%%%%%%%%%%%%%%%%%%%%%%%%%
%%%%%%%%%%%% SECTION III %%%%%%%%%%%%%
%%%%%%%%%%%%%%%%%%%%%%%%%%%%%%%%%%%%%%
\section{The Proposed Delay Structure and APUF Design}
\label{sec:proposed}
Before going over the architecture of the proposed design, we will first lay out some preliminary assumptions and FPGA primitives we are utilizing.
Most FPGA vendors such as Altera or Xilinx partition their resources into slices.
Typically, slices for most FPGA architectures of these FPGA vendors consist of four Look-Up Tables (LUTs), a Carry-Chain (CC), four Flip-Flops (FFs) and other resources. 
In our design, we utilize these slice primitves thoroughly.
Our design strongly utilizes the slice layout on Xilinx FPGA architectures, however this design can be mapped to other vendor architectures since almost all vendors include the primitives of LUTs, CCs, and FFs in their slice.

%%%%%%%%%%%%%%%%%%%%%%%%%%%%%%%%%%%%%%%%%
%%%%%%%%%%% FIGURE 3 %%%%%%%%%%%%%%%%%%%%
%%%%%%%%%%%%%%%%%%%%%%%%%%%%%%%%%%%%%%%%%
\begin{figure*}[t]
\centering
\includegraphics[width=\textwidth]{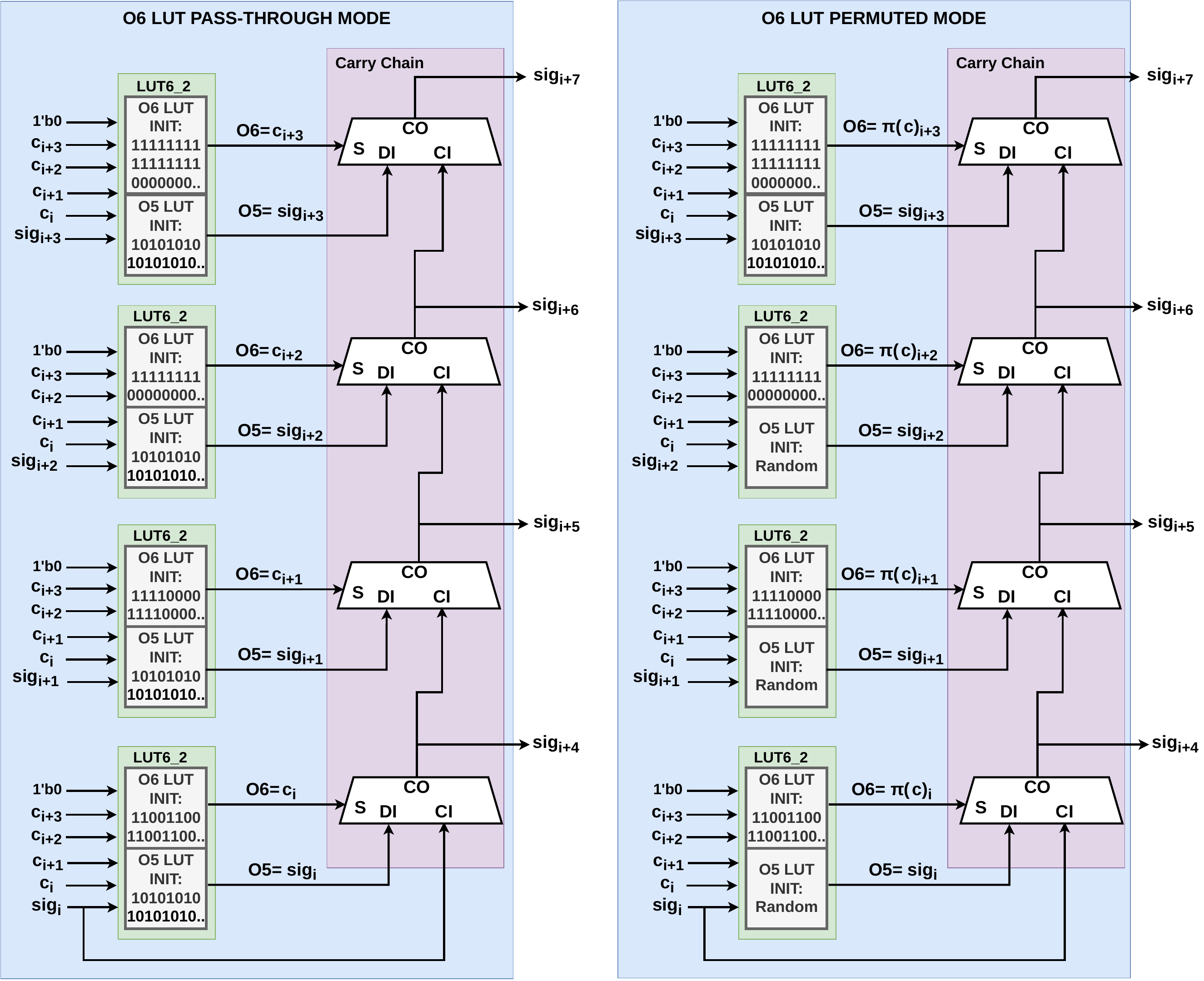}
\caption{Our Slice-based Delay Structure using LUTs and Carry-Chain.}
\Description{Slice-based Delay Strutcure is shown utilizing the four LUTs in the slice and the carry chain. Challenge bits go through the LUT to the S input of the carry chain. The other output of the LUT goes to the DI input of the carry chain.}
\label{fig:novel_delay_block}
\end{figure*}
%%%%%%%%%%%%%%%%%%%%%%%%%%%%%%%%%%%%%%%%%
%%%%%%%%%%%%%%%%%%%%%%%%%%%%%%%%%%%%%%%%%
%%%%%%%%%%%%%%%%%%%%%%%%%%%%%%%%%%%%%%%%%

\subsection{The Novel Delay Stage Design}
As~\autoref{fig:novel_delay_block} shows, our delay stage uses a complete
typical FPGA slice. It has four inputs from the previous stage $\vec sig = [sig_{i+3}, sig_{i+2}, sig_{i+1}, sig_{i}]$, four challenge-bit inputs $\vec c = [c_{i+3}, c_{i+2}, c_{i+1}, c_{i}]$, and four outputs going to the next stage $\vec osig = [sig_{i+7}, sig_{i+6}, sig_{i+5}, sig_{i+4}]$. Within the slice, each dual-output 6-input LUT primitive, LUT6\_2, is used as two 5-input LUT functions with outputs O5 and O6. This is done by tying input I5 of the LUT6\_2 to logic 1. 
The O5 output is hardwired to the data input, DI, of the corresponding multiplexer in the carry chain, while the O6 output is hardwired to the select input, S, of that
multiplexer~\cite{7seriesxilinxclbguide}. 
This LUT-to-carry-chain connection is fixed in the slice and can not be changed.
For each of the four LUT6\_2 primitives, I0 receives the corresponding incoming chain signal, while I1--I4 receive the same four challenge bits. 
Input I5 is tied to logic 1. 
The O5 function is initialized such that it passes the incoming chain signal through the LUT:
\begin{equation}
    O5_j = sig_j,
    \qquad j \in \{0,1,2,3\},
    \label{eq:o5_passthrough}
\end{equation}
where $sig_j$ is the corresponding incoming chain signal. This behavior is implemented by setting the lower 32 bits of the LUT initialization value to \texttt{32'hAAAA\_AAAA}, which produces $O5=I0$.

We can now focus on the four multiplexers in the carry chain. For each multiplexer apart from the first, one input is the carry output of the preceding multiplexer, while the other input is provided through the O5 output of the corresponding LUT6\_2. 
This connection is also fixed by hardwiring in the FPGA slice and can not be changed.
Further, in our implementation, the first incoming chain signal, $sig_0$, is also connected to the CI input of the first mux in the CARRY4 primitive, as shown in \autoref{fig:novel_delay_block}. 
The four O6 outputs of the four LUT6\_2 primitives in the slice provide the select signals $S_0$--$S_3$ of the four carry-chain multiplexers.
The delay block can be configured in two different O6 modes. 
In both modes, the O5 behavior remains unchanged and continues to pass the four incoming chain signals to the DI inputs of the carry chain. 
The two modes differ only in how the O6 functions map the 4-bit challenge to the four carry-chain select signals.
Before we go into the modes, we will first explain how we use the O6 function of the four LUTs to map the four challenges to the select lines of the four muxes.
Since each LUT6\_2 in the slice gets the same four challenge inputs and the O6 wire is fixed to the select bit of the corresponding mux:
\begin{equation}
    S_{j}
    = O6_{j} = O6LUT_{j}(LUT\_INIT_{j}, \vec c),
    \qquad j \in \{0,1,2,3\},
    \label{eq:carrychain_select}
\end{equation}
where $S_j$ is the select bit of mux $j$ in the carry-chain, $LUT\_INIT_{j}$ is the initialization for the lookup table memory.
The LUT initialization and the block challenge vector $\vec c$ together decide the output O6 of the LUT6\_2 number $j$ in the slice.
Now we explain the two modes using this equation.
\paragraph{O6 Pass-Through Mode.}
In the pass-through mode, each LUT6\_2 is initialized such that its O6 output passes one of the four challenge bits directly to the select input of its corresponding carry-chain multiplexer. 
The first LUT in the slice passes challenge bit $c_i$ to the first carry-chain multiplexer, the second LUT passes $c_{i+1}$ to the second multiplexer, and so on. 
Therefore,
\begin{equation}
    [S_3,S_2,S_1,S_0]
    =
    [c_{i+3},c_{i+2},c_{i+1},c_i].
    \label{eq:o6_passthrough}
\end{equation}
Equivalently, if
\begin{equation}
    \vec{c} =
    [c_{i+3},c_{i+2},c_{i+1},c_i],
    \label{eq:block_challenge}
\end{equation}
then
\begin{equation}
    \vec{S}=\vec{c}.
    \label{eq:passthrough_final}
\end{equation}
This behavior is our non-permuted LUT6\_2 initialization where the 4-bit challenge value is used directly as the carry-chain select state.
Essentially $LUT\_INIT_{j}$ for the four LUT6\_2 are chosen such that~\eqref{eq:carrychain_select} becomes equivalent to~\eqref{eq:passthrough_final}. 

\paragraph{O6 Permutation Mode.}
In the permutation mode, the O5 functionality remains unchanged, but the four O6 functions are initialized jointly to implement a randomized mapping between the 4-bit challenge and the 4-bit carry-chain select state. 
The resulting select vector is
\begin{equation}
    \vec{S}=\pi(\vec{c}),
    \label{eq:o6_randomized}
\end{equation}
where $\pi$ is a seed-generated permutation over the 16 possible 4-bit challenge states.
There are $2^4=16$ possible values of $\vec{c}$ and also 16 possible configurations of $\vec S$. 
The permutation mode assigns every challenge state to exactly one select state, with every select state occurring exactly once. 
The four LUTs jointly implement one permutation of the complete 4-bit select space of $\vec S$. 
Since there are 16 possible states, there are $16!$ possible permutations.
Algorithm~\ref{alg:lut_init_gen} shows how LUT\_INIT for each LUT in the slice can be generated based on mode and permutation provided.
A verilog function can be designed to take as input the parameters and generate the respective LUT\_INIT  for each LUT in the slice.
\autoref{tab:ctosmapping} shows how $\vec c$ maps to $\vec S$ in each mode.
For the permutation mode, the table shows an example for one particular permutation.
As discussed before, $16!$ permutations are possible. 

\begin{table}[t]
\centering
\caption{$\vec c$ to $\vec S$ mapping in both Modes}
\label{tab:ctosmapping}
\begin{tabular}{lll}
\hline
\textbf{$\vec c$} & \textbf{$\vec S$ Pass-Through Mode} & \textbf{$\vec S$ Permuted Mode (One Example Permutation)} \\
\hline
\texttt{4'b0000} & \texttt{4'b0000} & \texttt{4'b0010} \\
\texttt{4'b0001} & \texttt{4'b0001} & \texttt{4'b1101} \\
\texttt{4'b0010} & \texttt{4'b0010} & \texttt{4'b1001} \\
\texttt{...} &  &  \\
\texttt{4'b1111} & \texttt{4'b1111} & \texttt{4'b0111} \\
\hline
\end{tabular}
\end{table}

\begin{algorithm}[t]
\footnotesize
\caption{LUT INIT Generator}
\label{alg:lut_init_gen}

\KwIn{\\
    \texttt{LUT Number,} \textbf{\texttt{j}};\\ 
    \texttt{Init Mode,} \textbf{\texttt{m}};\\
    \texttt{Permutation for $\vec c \to \vec S$,} \textbf{\texttt{p}};
}
\KwOut{\\
\texttt{LUT j initialization parameter,} \textbf{\texttt{LUT\_INIT}}
}

\BlankLine
\texttt{LUT\_INIT} $\gets$ \texttt{0xAAAA\_AAAA} \tcp*[r]{lower 32 bits of INIT set to pass-through I0 or $sig_{j}$ to O5}
\If{\texttt{m == PASS\_THROUGH}}{
\Switch{\texttt{j}}{
    \Case{$0$}{
        \texttt{LUT\_INIT} $\gets$ \texttt{0xCCCC\_CCCC} $||$ \texttt{LUT\_INIT}\tcp*[r]{$c_{0}$ passed through}
    }
    \Case{$1$}{
        \texttt{LUT\_INIT} $\gets$ \texttt{0xF0F0\_F0F0} $||$ \texttt{LUT\_INIT}\tcp*[r]{$c_{1}$ passed through}
    }
    \Case{$2$}{
        \texttt{LUT\_INIT} $\gets$ \texttt{0xFF00\_FF00} $||$ \texttt{LUT\_INIT}\tcp*[r]{$c_{2}$ passed through}
    }
    \Case{$3$}{
        \texttt{LUT\_INIT} $\gets$ \texttt{0xFFFF\_0000} $||$ \texttt{LUT\_INIT}\tcp*[r]{$c_{3}$ passed through}
    }
    \Other{
        \texttt{LUT\_INIT} $\gets 0$\tcp*[r]{Incorrect input}
    }
}
}
\ElseIf{\texttt{m == PERMUTE}}{
\For{\texttt{i} $\gets 0$ \texttt{to} $15$}{
    \texttt{combination[i]} $\gets$ \texttt{hex(i)}\tcp*[r]{create hex vector of all 4 bit combinations}
}
\tcc{permutation algorithm is left to the designer. Seed can specify one of the 16! possible permutations. We implement a fisher-yates seeded shuffler in our design.}
\texttt{combination} $\gets$ \texttt{permute(seed, combination)}\;

% \texttt{UPPER\_LUT\_INIT} $\gets$ 0 \tcp*[r]{Upper 32 bits of init initialized to 0}
\tcc{assign LUT j with the jth bit of 4-bit output against i for this particular permutation and repeat for 16 elements. Repeat the same bit twice in the init since I0 is for $sig_j$, so we get same output regardless of whether IO is 0 or 1.}
\For{\texttt{i} $\gets 0$ \texttt{to} $15$}{
     \texttt{LUT\_INIT} $\gets$ \texttt{combination[i][j]} $||$ \texttt{combination[i][j]} $||$ \texttt{LUT\_INIT}\;
}

}

\BlankLine
\Return{\texttt{LUT\_INIT}}\;

\end{algorithm}

\subsection{HAPUF Architecture Design}
Now that we have explained our delay block design, we present our proposed HAPUF architecture consisting of trigger flops, two parallel delay lines followed by the final arbiter(s). 
Each delay line is constructed from a sequence of the proposed carry-chain based delay blocks. 
The two delay lines use the same number and ordering of delay blocks, and corresponding blocks are implemented using equivalent FPGA resources.
Each delay line begins with a dedicated line trigger flop. 
A primary trigger flop first generates the transition used for the PUF evaluation, and its output is then provided to two line-trigger flip-flops, one for each delay line. Following~\cite{sahoo2015towardsidealapuf}, one complete FPGA slice is reserved for the primary-trigger flip-flop and one
complete slice is reserved for each line-trigger flip-flop. 
This is done to reduce systematic bias introduced by asymmetric FPGA routing between the trigger circuitry and the two delay lines.

The output of each line trigger provides the initial transition to its corresponding delay line. 
From this point onward, the transition propagates through the sequence of carry-chain delay blocks until the final output of each delay line reaches the arbiter.
The two delay lines never cross each other. 
This is intentional since crossing between the two lines can introduce additional routing asymmetry from the FPGA row and column interconnect structure. 
Instead, both lines remain physically separate, while the placement of corresponding delay blocks is constrained to be as symmetric as possible.
It should be noted that since the delay blocks in each line has four parallel signals propagating through the line, there is some routing asymmetry being introduced, however, our delay lines never cross into another SLICE column to avoid major routing asymmetries as in crossing implementations with different slice columns for delay lines as discussed in~\autoref{subsection:background}.

Ideally, if the deterministic delay contribution of both lines were identical, the final delay difference would be determined primarily by device-specific process variation between corresponding FPGA resources. 
In practice, perfect routing symmetry is difficult to achieve because nominally equivalent placements can still use different routing resources. 
The tuning blocks described later are therefore used to compensate for systematic delay imbalance between the two lines.

%%%%%%%%%%%%%%%%%%%%%%%%%%%%%%%%%%%%%%%%%
%%%%%%%%%%% FIGURE 4 %%%%%%%%%%%%%%%%%%%%
%%%%%%%%%%%%%%%%%%%%%%%%%%%%%%%%%%%%%%%%%
\begin{figure*}[t]
\centering
\includegraphics[width=\textwidth]{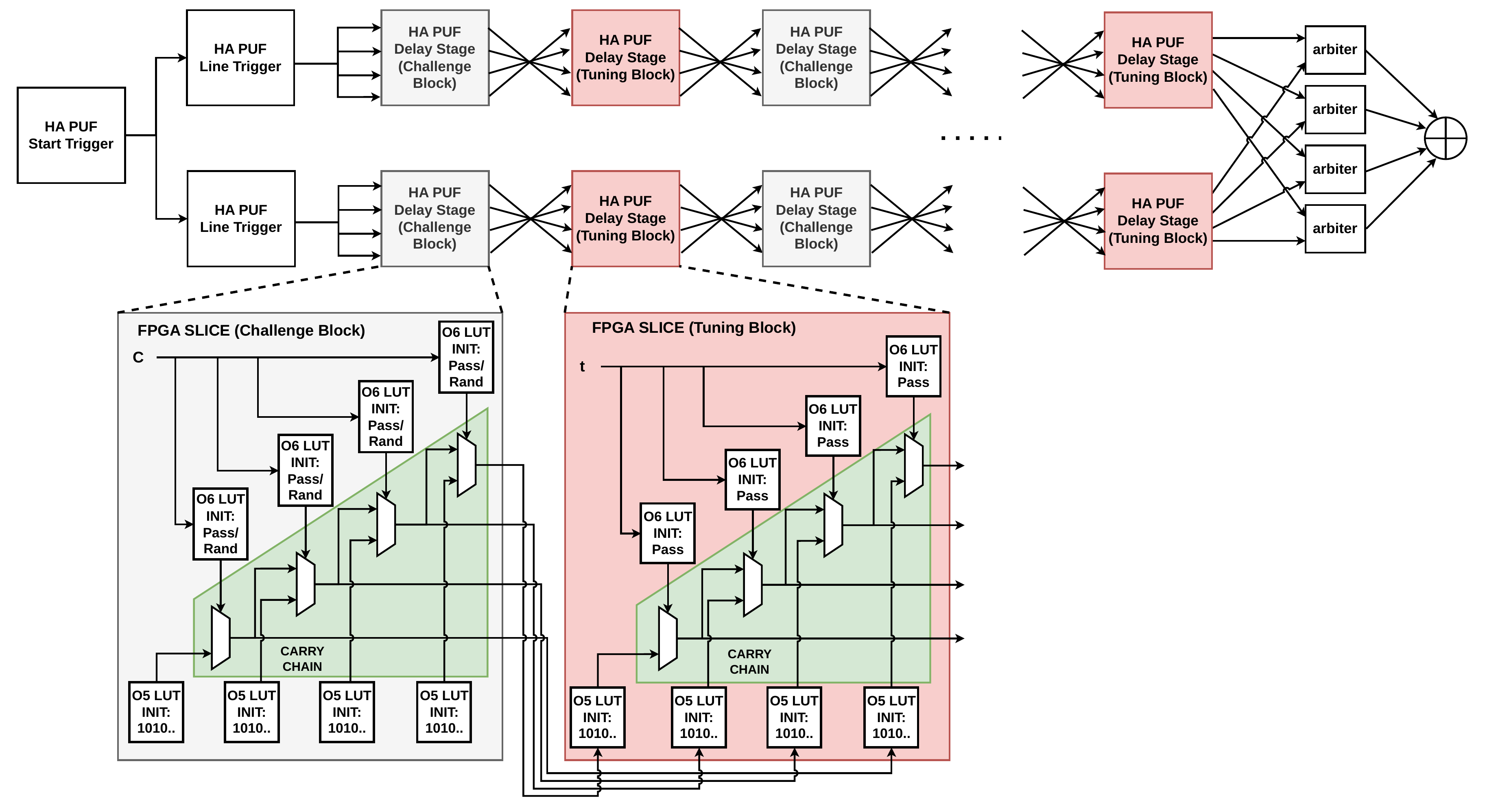}
\caption{Helix Arbiter PUF delay stage onto FPGA slice.}
\Description{Helix Arbiter PUF architecture is shown. A start trigger flop is followed by two line flops. Each line contains the delay stage block placed consecutively. These blocks act as challenge blocks and tuning blocks alternately. The output of the final delay block goes to four arbiters. The four arbiter results are XOR's to give a single response.}
\label{fig:HAPUF_arch}
\end{figure*}
%%%%%%%%%%%%%%%%%%%%%%%%%%%%%%%%%%%%%%%%%
%%%%%%%%%%%%%%%%%%%%%%%%%%%%%%%%%%%%%%%%%
%%%%%%%%%%%%%%%%%%%%%%%%%%%%%%%%%%%%%%%%%

\subsection{HAPUF Delay-Line Construction}

Each delay line is constructed from a sequence of the delay blocks described previously. 
Every delay block receives four signals from the preceding block and produces four signals for the following block.
The four outputs of one block are connected to the next block in a
mirrored fashion, as shown in \autoref{fig:HAPUF_arch}. 
In particular, the external data input of mux~0 in block $g$ is driven by mux~3 of block $g-1$, the input of mux~1 is driven by mux~2, the input of mux~2 is driven by mux~1, and the input of mux~3 is driven by mux~0.

Let $j\in\{0,1\}$ index the delay line, $g=1,\dots,G$ the block, and $k\in\{0,1,2,3\}$ the four muxes in the delay block, and therefore let $\Delta_{j,g,k}$ represent the accumulated propagation delay at mux $k$ of delay block $g$ in delay line $j$. 
The four externally supplied signals to delay block $g$ therefore correspond to
\begin{equation}
    \left[
    \Delta_{j,g-1,3},
    \Delta_{j,g-1,2},
    \Delta_{j,g-1,1},
    \Delta_{j,g-1,0}
    \right].
    \label{eq:hapuf_mirror_inputs}
\end{equation}
And let
\begin{equation}
    \vec{S}_g =
    [S_{g,0},S_{g,1},S_{g,2},S_{g,3}]
\end{equation}
denote the four effective select signals applied to the carry-chain multiplexers of delay block $g$.
For the pass-through mode, we can drop the delay-line specifier $j$ here since both lines get the same challenge bits or carry-chain select bits $\vec S$ on account of O6 being pass-through.
We can also drop $j$ and assume identical $\vec S$ for the permuted mode in the specific case when both lines are supplied identical seeds.
With identical seeds, even though $\vec c \to \vec S$ are permuted per delay blocks, the identical seed makes sure they are deterministically identical permutations for both delay lines.
With permuted mode with different seeds for each delay line, the specifier $j$ becomes relevant in $S_{j,g,k}$ since each block has a unique permutation $\vec c \to \vec S$.
We continue here for now with the pass-through or same seed for both lines case.

For mux~0, both the DI input and the carry input originate from
$\Delta_{j,g-1,3}$. However, the two signals reach the carry-chain
multiplexer through different physical paths. The delay at the output
of mux~0 can therefore be represented as
\begin{equation}
    \Delta_{j, g,0}
    =
    \Delta_{j,g-1,3}
    +
    (1-S)\alpha_{j,g,0}
    +
    S\beta_{j,g,0}.
    \label{eq:hapuf_mux0}
\end{equation}
Here, $\alpha_{j,g,0}$ and $\beta_{j,g,0}$ represent the delays associated with the two physical routes selected by the first carry-chain multiplexer.
It can be visualized that one route is an input to the slice directly to the carry-chain while the other is through the pass-through O5 LUT.

For mux 1 -- 3, each mux takes as input the mirrored output of the last delay block $\Delta_{j,g-1,3-k}$, and the output of the previous mux in the carry-chain of the same delay block $\Delta_{j,g,k-1}$ :
\begin{align}
  \Delta_{j,g,k}
  &= (1-S_{g,k})\left[\Delta_{j,g,k-1}+\alpha_{j,g,k}\right]
  \nonumber\\
  &\quad + S_{g,k}\left[\Delta_{j,g-1,3-k}+\beta_{j,g,k}\right],
  \qquad k=1,2,3.
  \label{eq:delay_recurrence}
\end{align}
For $k = 1,2,3$, we can consider $\alpha_{j,g,k}$ to represent the wire delay between carry-chain muxes in the same delay block, while $\beta_{j,g,k}$ represents the wire delay from the mirrored output from the previous block, through the O5 LUT and into the carry-chain mux.
We now attempt to recursively expand the four outputs of a group $\Delta_{j, g, k}$.
We represent the delay accumulated at $\Delta_{j, g, 1}$ by plugging in $\Delta_{j, g, 0}$ from~\autoref{eq:hapuf_mux0}: 
\begin{align}
  \Delta_{j,g,1}
  &= (1-S_{g,1})\left[\Delta_{j,g-1,3}+(1 -S_{g,0})\alpha_{j,g, 0} + S_{g,0}\beta_{j,g,0} +\alpha_{j,g,1}\right]
  \nonumber\\
  &\quad + S_{g,1}\left[\Delta_{j,g-1,2}+\beta_{j,g,1}\right].
  \label{eq:k1-step}
\end{align}
Next we attempt expand for $\Delta_{j,g,2}$ by plugging in $\Delta_{j,g,1}$:
\begin{align}
  \Delta_{j,g,2}
  &= (1-S_{g,2})\{(1-S_{g,1})[\Delta_{j,g-1,3}+(1 -S_{g,0})\alpha_{j,g, 0} + S_{g,0}\beta_{j,g,0} +\alpha_{j,g,1}]
  \nonumber\\
  &\quad + S_{g,1}(\Delta_{j,g-1,2}+\beta_{j,g,1}) +\alpha_{j,g,2}\}
  \nonumber\\
  &\quad + S_{g,2}\left[\Delta_{j,g-1,1}+\beta_{j,g,2}\right].
  \label{eq:k2-step}
\end{align}
We can continue in this way to expand for $\Delta_{j,g,3}$.
We can see that this recurrence will continue to grow block by block.
The important takeaway here is that as the line delay accumulates, the wire delays $\alpha_{j,g,k}$ and $\beta_{j,g,k}$ are part of the challenge selected delays and do not grow independent of challenge as bias in the delay lines.
This is an improvement over the non-crossing standard APUF model in~\autoref{subsection:background} where bias grows with stages and challenge controlled delay is insufficient to affect response as shown in~\eqref{eq:delta_weight}.
On the other hand, with the path-crossing standard APUF implementation in ~\autoref{subsection:background} the two lines exchange signals, so the crossing inverts the sign of the delay difference as shown in~\autoref{eq:apuf-recurrence}.
As the signal alternates between two physical tracks, any systematic difference between those tracks is traversed on every stage where the signal is on the slower one, so it is accumulated into every $w_i$ rather than appearing once. 
On FPGAs, where track asymmetry is set by the routing fabric and is typically larger than the process variation being measured, this systematic component dominates the random variations.
Which affects the uniqueness.
Our design avoids the large asymmetry by avoiding crossing SLICE columns which introduces significant asymmetry.
We follow the principle of the non-crossing implementation here that both delay lines follow the same path, so that the response is decided by random variation as muhc as possible.
We do this since both our delay lines are replicas of each other as much as possible, and get the same challenge or mux selection bits.
This attempts to make the route taken by the signals in both lines as similar as possible so that the response is most sensitive to the random variation in those identical paths.

\subsection{Challenge and Tuning Blocks}
Although every delay block uses the same physical FPGA structure, the blocks are used for two different purposes within the delay line. 
Some blocks are controlled directly by challenge bits, while others are controlled by tuning bits. 
We refer to these as \emph{challenge blocks} and \emph{tuning blocks}, respectively.
Let
\begin{equation}
    \vec{q}_g =
    [q_{g,0},q_{g,1},q_{g,2},q_{g,3}]
\end{equation}
represent the 4-bit control word supplied to delay block $g$. 
For a challenge block, this control word is taken from the external PUF challenge, while for a tuning block it is taken from the tuning vector:
\begin{equation}
    \vec{q}_g =
    \begin{cases}
        \vec{c}_g, & \text{if block $g$ is a challenge block},\\[4pt]
        \vec{t}_g, & \text{if block $g$ is a tuning block}.
    \end{cases}
    \label{eq:hapuf_control_word}
\end{equation}
The challenge bits determine the challenge-dependent propagation path through the delay line in each evaluation as they change between evaluations.
The tuning bits, in contrast, are selected through software
and can be held constant while a set of challenges is evaluated.
In our implementation, challenge blocks and tuning blocks are
distributed throughout the delay line in an alternating arrangement.
This places tuning capability at multiple positions along the propagation path rather than concentrating it at the end of the delay line. 
As a result, tuning can modify the accumulated delay at several points before the final arbiter comparison.
In our alternating arrangement, tuning blocks are placed between challenge blocks.
Therefore we can imagine that tuning blocks tune the delays $\beta_{j,g,k}$ for $k= {0,1,2,3}$ and $\alpha_{j,g,k}$ for $k= {0}$ in~\eqref{eq:hapuf_mux0} and~\eqref{eq:delay_recurrence}. 
For $k = {1,2,3}$, $\alpha_{j,g,k}$ is just the delay between carry-chain muxes within the delay-block slice so it can not be affected by the tuning blocks.

Since both the challenge and tuning vectors are supplied through software, their corresponding bit ranges can be assigned independently to the appropriate delay blocks without changing the physical architecture of the delay line.

\subsection{Delay-Block Operating Modes}
The same physical delay block can be configured in either O6 pass-through mode or O6 permuted mode. 
The O5 functionality remains unchanged in both cases and continues to pass the incoming chain signals to the DI inputs of the carry chain. 
The only difference is the mapping between the 4-bit block control word and the effective carry-chain select vector.\newline
In O6 pass-through mode, the control bits are passed directly to the four carry-chain multiplexers:
\begin{equation}
    \vec{S}_g = \vec{q}_g.
    \label{eq:hapuf_pass_mode}
\end{equation}
Therefore,
\begin{equation}
    \vec{S}_g =
    \begin{cases}
        \vec{c}_g, & \text{for a challenge block},\\[4pt]
        \vec{t}_g, & \text{for a tuning block}.
    \end{cases}
\end{equation}
In this mode, corresponding blocks in the two delay lines receive the same challenge or tuning control word and therefore use the same logical carry-chain select state. 
Any difference in the resulting propagation delay is then caused by the physical delay differences between the corresponding FPGA resources.

In O6 permuted mode, the four LUT6\_2 primitives in each delay block jointly implement a fixed permutation of the 4-bit carry-chain mux select space. 
In permuted mode, the alternating delay-blocks we choose to be tuning blocks are kept in pass-through mode.
Challenge-blocks are instantiated in their permuted mode.
Since the permutation is specific to both the delay line and the position of the challenge block within that delay line, we redefine $\vec S_g$ in permuted mode and write:
\begin{equation}
    \vec{S}_g =
    \begin{cases}
        \pi_{g}(\vec{c}_g), & \text{for a challenge block},\\[4pt]
        \vec{t}_g, & \text{for a tuning block}.
    \end{cases}
\end{equation}
More precisely for challenge-blocks in delay line $j$ we say:
\begin{equation}
    \vec{S}_{j,g}
    =
    \pi_{j,g}(\vec{c}_{j,g}),
    \label{eq:hapuf_random_mode}
\end{equation}
where $\pi_{j,g}$ is the 4-bit permutation implemented by delay block $g$ in delay line $j$.
Each delay line is assigned its own seed, denoted by $\nu_j$. 
The permutation supplied to each individual delay block is then deterministically derived from the seed of its delay line $\nu_j$ and the position of the block $g$ within that line. 
The permutation for delay block $g$ in delay line $j$ can therefore be written as
\begin{equation}
    \pi_{j,g}
    =
    f(\nu_j,g),
    \label{eq:hapuf_block_seed}
\end{equation}
where $f(\cdot)$ is the deterministic permutation-generation function used during LUT initialization.
The block permutation $\pi_{j,g}$ determines the permutation for the particular delay block. 
Therefore, different blocks within the same delay line can implement different challenge-to-select mappings, even though their physical delay-block structures are identical. 
With regards to the line seed, there are two options here.
The two lines may use the same seed:
\begin{equation}
    \nu_0 = \nu_1.
    \label{eq:hapuf_line_seeds_diff}
\end{equation}
Or the two delay lines may use different seeds:
\begin{equation}
    \nu_0 \neq \nu_1.
    \label{eq:hapuf_line_seeds}
\end{equation}
In the case of same seeds, $\vec S_{0,g} = \vec S_{1,g}$ which is the same as the pass-through implementation.
The only difference is that $\vec c_{j,g} \to \vec S_{j,g}$ or $\pi_{j,g}$ is unique per block, but the same for the same block positions in the same line $\pi_{0,g} = \pi_{1,g}$. But, in the case of a secret seed, it can serve as a per-block challenge obfuscation method.
% The permutation supplied to each individual delay block is then deterministically derived from the seed of its delay line and the position of the block within that line. 
% The permutation for delay block $g$ in delay line $j$ can therefore be written as
% \begin{equation}
%     \pi_{j,g}
%     =
%     f(\nu_j,g),
%     \label{eq:hapuf_block_seed}
% \end{equation}
In the case of different seeds for each delay line, blocks in the same position $g$ in the two lines generally implement different permutations:
\begin{equation}
    \pi_{0,g} \neq \pi_{1,g}.
    \label{eq:hapuf_different_permutations}
\end{equation}
Consequently, applying the same external 4-bit challenge to corresponding challenge blocks in the two delay lines does not necessarily result in the same internal carry-chain select state. 
For the two delay lines,
\begin{equation}
    \vec{S}_{0,g}
    =
    \pi_{0,g}(\vec{c}_{0,g}),
    \neq
    \vec{S}_{1,g}
    =
    \pi_{1,g}(\vec{c}_{1,g}).
    \label{eq:hapuf_line_select_vectors}
\end{equation}
For corresponding challenge blocks, the same challenge bits can therefore select different internal carry-chain paths in the two delay lines. 
Corresponding tuning blocks on the other hand are pass-through since their function is for tuning the delay line for better metrics.
\begin{equation}
    \vec{s}_{j,g}
    =
    \vec{t}_g
    \label{eq:hapuf_random_tuning}
\end{equation}
Thus, the two delay lines retain the same number of blocks, the same block ordering, and the same overall physical architecture, while their internal path selections can differ because the O6 mappings are generated from different line seeds.
The physical delay recurrence in \eqref{eq:delay_recurrence} for the case of different seeds for each delay line changes so the the line specific select vectors $\vec S_{j}$ are different, so both lines choose different paths.
As we will see in~\autoref{sec:methodology}, because of this change, our instance becomes unique to the unique different-seed pair used.

\subsection{HAPUF Arbiter}

After propagation through the complete sequence of delay blocks, the final output of each delay line is provided to flip-flop-based arbiter. 
One complete FPGA slice is reserved for each arbiter flip-flop.
Let $G$ denote the final delay block in each delay line. 
The final accumulated delays on each of the four outputs of the final delay block in each line are:
\begin{equation}
    [\Delta_{0,G,0}, \Delta_{0,G,1}, \Delta_{0,G,2}\Delta_{0,G,3}]
    \qquad \text{and} \qquad
    [\Delta_{1,G,0}, \Delta_{1,G,1}, \Delta_{1,G,2}\Delta_{1,G,3}]
\end{equation}
To counteract the affect of routing asymmetry on uniqueness in our pass-through mode, we use four arbiter flops for each pair of the four outputs of the final delay block.
The HAPUF response is determined by then XORing the results of the four arbiters:
\begin{equation}
    \rho_{i}= 
    \begin{cases}
        1\text{ if }\Delta_{0, G, i} > \Delta_{1, G, i} \\
        0\text{ if }\Delta_{0, G, i} \leq \Delta_{1, G, i}
    \end{cases}
    \qquad i = 0,1,2,3
    \label{eq:stdapuf_resp}
\end{equation}
where $\rho_{i}$ is the result of arbiter flop number $i$.
And:
\begin{equation}
    R = \rho_0 \oplus \rho_1 \oplus\rho_2 \oplus \rho_3
    \label{eq:hapuf_response}
\end{equation}
Thus, the final response depends on the accumulated propagation delays produced by the complete sequence of challenge-controlled and tuning-controlled delay blocks in the two parallel delay lines. 

In permuted mode with different seeds $\nu_0 \neq \nu_1$, since both delay lines have a different permutation for $\vec c \to \vec S$ in each challenge block, the different permutation seed used in each block determines the response rather than inter-device variations.
Therefore our permuted mode PUF instance is more line-seeds based rather than device-based.
We deviate from the PUF definition here since since PUFs are unique across device, but in our permuted mode, this "PUF" instance is unique across line-seeds as we will show in experimentation.
We make the design decision to make the line-seeds secret, and since the routing difference between the delay lines against the same challenge on account of different permutation seeds determines the response and uniqueness, we only make use of one of the signals output from the final blocks in the delay lines and a single arbiter:
\begin{equation}
    R = \rho_3
\end{equation}
We will show in~\autoref{sec:methodology}, that our pass-through mode PUF is unique across devices, while our permuted more PUF is unique across unique line-seeds across multiple devices and non-unique across devices for the same line-seeds.

\subsection{HAPUF Tuning}
As discussed, some of the blocks in the delay lines are reserved for tuning.
We find it preferable to keep the the control words supplied to the delay blocks $\vec q$ to be kept software controllable through registers which makes it possible to decide on the fly which blocks are tuning blocks and which are challenge blocks.
Control words for the tuning blocks are kept fixed once a suitable tuning has been found while the control words for the challenge blocks are then varied with the supplied challenge.
We prefer to keep tuning blocks alternatively as it allows bias to be countered at every stage, although other configurations may be more suitable to other implementations.
Tuning blocks were first introduced by Majzoobi et al.~\cite{majzoobi2010fpga}, who demonstrate them placed in the end but also suggested placing them anywhere the designer wishes.
Using these delay blocks, we now present our algorithm which we use to tune our PUF instance for metrics such as uniformity and neighbour-difference which measures how often responses for neighbouring challenges are the same.
We introduce some notation before explaining the algorithm.
Let $T_0$ and $T_1$ denote the complete tuning words applied to delay lines 0 and 1, respectively.
For the $i$th challenge $C_i$, let $\rho_{i,j}$ denote the response of arbiter $j$ and
let $R_i$ denote the final PUF response.
For the pass-through implementation, the response is $R_i = \rho_{i,0}\oplus \rho_{i,1}\oplus \rho_{i,2}\oplus \rho_{i,3}$ while for the permuted implementation it is just $R_i = \rho_{i,3}$ as established before.
The algorithm covers only the pass-through case since both are the same apart from the final response, but for a permuted implementation, only a single arbiter can be considered in the algorithm.
$U_{\oplus}$ denotes the uniformity of the final response, $ND_{\oplus}$ denotes the neighbor difference which is a check if the current response $R_i$ and the last response $R_{i-1}$ was the same , $U_j$ denotes the uniformity of arbiter $j$, and $HD_{j,k}$ denotes the Hamming distance between the responses of arbiters $j$ and $k$.
For the permuted variant, we would only consider metrics of one arbiter.
Values close to $0.5$ are desired for each of these metrics.
$\sigma$ represents the score of the current tuning combination and $\sigma_{min}$ is the best or lowest tuning score found so far.
$P_U$ abd $P_{HD}$ represents the penalties applied to the score for each tuning run.

Algorithm~\ref{alg:puf_tuning} presents the sequential seach form of our tuning procedure.
The algorithm first initializes the minimum score $\sigma_{min}$ and the best tuning pair $T^*_0$ and $ T^*_1$ (line~\ref{alg:init}).
The algorithm then creates nested loops to try every possible combination of the complete tuning words $T_0$ and $T_1$ for the two delay lines (line~\ref{alg:tuneloops}). 

Next, the algorithm initializes the statistics such as number of ones in the response $xorOnes$, number of ones in individual arbiters $arbOnes[j]$, and neighbours difference in response $neighbourChanges$, and consecutive runs of 1s or 0s in $runLen$ which are used to evaluate each candidate (lines~\ref{alg:counters1}--\ref{alg:counters2}).
For every tuning pair, the algorithm generates $N$ deterministic pseudorandom challenges using SplitMix64 (line~\ref{alg:challenge}).
SplitMix64 is used so that a significant distribution of challenge bits can be used rather than just the first N challenges from a very large space.
Further, FPGA APUFs generally have poor neighbor difference metrics due to linear design.
SplitMix64 provides some pre-processing to counter this. 
The same challenge sequence is used for every candidate, ensuring that tuning configurations are compared under identical input conditions.
The algorithm then applies the challenge and tuning words to the PUF and obtains the four individual arbiter responses (line~\ref{alg:challenge}).
Their XOR is then calculated and constitutes the final PUF response (line~\ref{alg:xor}).
For the permuted implementation, we would only consider one of the arbiters $\rho_3$ as the response here.

The number of ones in the final response and in each individual arbiter response are then accumulated, as well as the pairwise disagreements between arbiters (lines~\ref{alg:uniformity}--\ref{alg:pairhd}).
This is to help in calculating uniformity and inter-HD between arbiters later for scoring.
Additionally changes between consecutive responses are counted and the length of consecutive identical responses are tracked (line~\ref{alg:neighbor}).
Long sequences of either zeros or ones are explicitly detected.
If this run length exceeds the threshold $R_{\max}$, the candidate is rejected early rather than spending additional evaluations on what must be a biased tuning (line~\ref{alg:runreject}).
Choice of $R_{\max}$ is left up to the designer for the strictness in metric needed.
Our choice was $R_{\max} = 50$.

For candidates that pass this test, we calculate the XOR-response uniformity $U_{\oplus}$, neighboring difference $ND_{\oplus}$, individual-arbiter uniformities $U_j$, and pairwise inter-arbiter Hamming distances $HD_{j,k}$ (lines~\ref{alg:xormetrics}-~\ref{alg:arbmetrics}).
Values approaching $0.5$ are desirable for each of these statistics.
The algorithm therefore penalizes biased individual arbiters and strong similarity between arbiter responses (line~\ref{alg:penalties}).
An overall score is calculated that primarily minimizes the deviations of XOR uniformity and neighboring difference from $0.5$, while $P_U$ and $P_{HD}$ act as secondary terms (line~\ref{alg:score}).
Whenever a lower score is obtained, the algorithm stores the corresponding tuning pair (line~\ref{alg:best}).
After all candidates have been evaluated, the algorithm returns the best tuning configuration (line~\ref{alg:return}).

We will clarify that for sufficiently large tuning bits and challenges evaluated, $N$, the number of iterations become too large to find a tuning in a reasonable amount of time.
We made use of a practical implementation where random combination of $T_0, T_1$ were tried until a viable pair was found.
This is more likely to find a solution faster since evenly spaced variation in the tuning bits are more likely to be a suitable tuning rather than the pairs encountered early on in a sequential search.
As most bits in $T_0$ and $T_1$ will be 0.
We present the sequential search nested for-loop exhaustive form in Algorithm~\ref{alg:puf_tuning} because the central contribution of the procedure is the tuning criterion itself rather than a particular optimization strategy used to traverse the tuning space.
The search strategy can therefore be adapted to the available computational resources without changing the underlying tuning objective.

\begin{algorithm}[t]
\footnotesize
\DontPrintSemicolon
\SetAlgoLined
\LinesNumbered

\KwIn{Tune width $W$ bits/line; number of challenges $N$}
\KwOut{Best tuning pair $(T_0^\star,T_1^\star)$}

$\sigma_{\min}\gets\infty,\quad
(T_0^\star,T_1^\star)\gets(0,0)$\;
\nllabel{alg:init}

\tcp{Examine every complete pair of tuning words}
\For{$T_0\gets0$ \KwTo $2^W-1$}{
    \For{$T_1\gets0$ \KwTo $2^W-1$}{
        \nllabel{alg:tuneloops}

        $xorOnes\gets0,\;
        neighborChanges\gets0,\;
        runLen\gets0$\;
        \nllabel{alg:counters1}

        $arbOnes[j]\gets0\ \forall j,\quad
        pairDiff[j,k]\gets0\ \forall j<k$\;
        $R_{\mathrm{prev}}\gets0,\quad valid\gets1$\;
        \nllabel{alg:counters2}

        \tcp{Test this tuning using the same pseudorandom challenges}
        \For{$i\gets0$ \KwTo $N-1$}{
            $C_i\gets
            \mathrm{SplitMix64}(i)\bmod 2^{n_c}$\;
            \nllabel{alg:challenge}

            $(\rho_{i,0},\rho_{i,1},\rho_{i,2},\rho_{i,3})
            \gets\mathrm{ReadPUF}(C_i,T_0,T_1)$\;
            \nllabel{alg:read}

            $R_i\gets
            \rho_{i,0}\oplus \rho_{i,1}\oplus
            \rho_{i,2}\oplus \rho_{i,3}$\;
            \nllabel{alg:xor}

            $xorOnes\gets xorOnes+R_i$\;
            $arbOnes[j]\gets arbOnes[j]+\rho_{i,j},
            \quad \forall j$\;
            \nllabel{alg:uniformity}

            $pairDiff[j,k]\gets pairDiff[j,k]
            +\mathbf{1}(\rho_{i,j}\neq \rho_{i,k}),
            \quad \forall j<k$\;
            \nllabel{alg:pairhd}

            \If{$i>0$}{
                $neighborChanges\gets neighborChanges
                +\mathbf{1}(R_i\neq R_{\mathrm{prev}})$\;

                $runLen\gets
                \begin{cases}
                    runLen+1, & R_i=R_{\mathrm{prev}}\\
                    1,       & R_i\neq R_{\mathrm{prev}}
                \end{cases}$\;
                \nllabel{alg:neighbor}
            }
            \Else{
                $runLen\gets1$\;
            }

            \tcp{Reject long runs of identical 0s or 1s}
            \If{$runLen>R_{\max}$}{
                $valid\gets0$; \textbf{break}\;
                \nllabel{alg:runreject}
            }

            $R_{\mathrm{prev}}\gets R_i$\;
        }

        \If{$valid=1$}{
            $U_{\oplus}\gets xorOnes/N,\quad
            ND_{\oplus}\gets neighborChanges/(N-1)$\;
            \nllabel{alg:xormetrics}

            $U_j\gets arbOnes[j]/N,\quad
            HD_{j,k}\gets pairDiff[j,k]/N$\;
            \nllabel{alg:arbmetrics}

            \tcp{Secondary penalties for biased/correlated arbiters}
            $P_U\gets
            \frac{1}{16}\sum_{j=0}^{3}
            \left|U_j-\frac{1}{2}\right|,\quad
            P_{HD}\gets
            \frac{1}{24}\sum_{j<k}
            \left|HD_{j,k}-\frac{1}{2}\right|$\;
            \nllabel{alg:penalties}

            \tcp{Lower score indicates better tuning}
            $\sigma\gets
            \left|U_{\oplus}-\frac{1}{2}\right|
            +
            \left|ND_{\oplus}-\frac{1}{2}\right|
            +P_U+P_{HD}$\;
            \nllabel{alg:score}

            \If{$\sigma<\sigma_{\min}$}{
                $\sigma_{\min}\gets S,\quad
                (T_0^\star,T_1^\star)\gets(T_0,T_1)$\;
                \nllabel{alg:best}
            }
        }
    }
}

\Return $(T_0^\star,T_1^\star)$\;
\nllabel{alg:return}

\caption{Tuning procedure for HAPUF.}
\label{alg:puf_tuning}
\end{algorithm}

%%%%%%%%%%%%%%%%%%%%%%%%%%%%%%%%%%
%%%%%%%%%%% SECTION IV %%%%%%%%%%%
%%%%%%%%%%%%%%%%%%%%%%%%%%%%%%%%%%
\section{Experimentation and Methodology}
\label{sec:methodology}
For implementing the proposed PUF, we will be using the Xilinx Arty A7-100t board (XC7A100T).
This choice is made due to the availability of the board and the Configurable Logic Block (CLB) architecture which consists of two slices with each slice consisting of four LUTs and a carry-chain that neatly maps onto our delay stage and max/min layer design~\cite{7seriesxilinxclbguide}.
Other boards are also suitable if the CLB and Slice layout are suitable for the delay stage.
Some FPGA architectures such as Ultrascale+ have eight LUTs and an eight-mux long carry-chain~\cite{ultrascalexilinxclbguide} per slice but each CLB consists of one slice.
In that case, the delay stages can be configured to be 8-inputs and 8-outputs wide for the slice-based delay block.
So the implementation is similar except our 4-bits parameterization moves to 8-bits.
\subsection{PUF implementation}
For the HAPUF implementation, we use Xilinx provided primitives such as `LUT6\_2', and `CARRY4' for implementing our delay blocks, and `FDCE' for triggers and arbiters.
The placement of these primitives are strictly controlled using `BEL' and `LOC' constraints.
Additionally `LOCK\_PINS' are also utilized to use specified pins so that routing to be as similar as possible.
The use of `FIXED\_ROUTE' constraints to make the routing in the two delay lines as identical as possible is encouraged but we were able to experiment without using them extensively and find suitable metrics through tuning. 
In the case of an absurdly routed wire causing a huge bias or a dominant challenge-bit effect `FIXED\_ROUTE' constraints may be needed.
Our pass-through implementation was tested with 7 challenge block and 7 tuning blocks alternately placed, while our different line-seed permuted implementation was tested with 5 challenge blocks and 5 tuning blocks alternately placed.
All code was written in verilog Hardware Description Language (HDL).

%%%%%%%%%%%%%%%%%%%%%%%%%%%%%%%%%%%%%%%%%
%%%%%%%%%%% FIGURE 3 %%%%%%%%%%%%%%%%%%%%
%%%%%%%%%%%%%%%%%%%%%%%%%%%%%%%%%%%%%%%%%
\begin{figure*}[t]
\centering
\includegraphics[width=\textwidth]{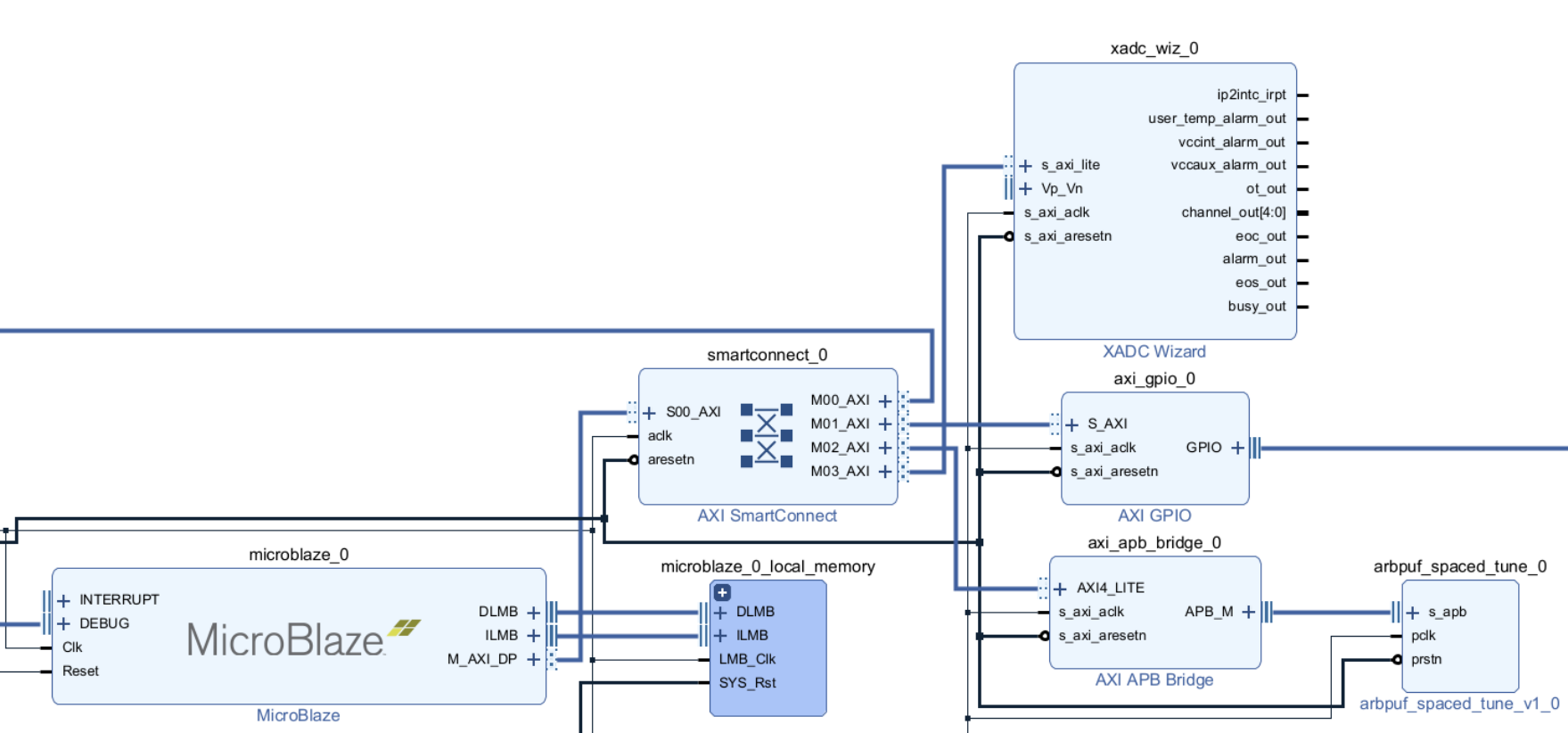}
\caption{Vivado block diagram of Microblaze Setup.}
\Description{A block design made using the Vivado tool is shown. It consists of a microblaze setup with AXI interconnects and APB bridges connnected to the PUF APB interface. An XADC wizard is also connected to microblaze via AXI.}
\label{fig:vitis_block_diagram}
\end{figure*}
%%%%%%%%%%%%%%%%%%%%%%%%%%%%%%%%%%%%%%%%%
%%%%%%%%%%%%%%%%%%%%%%%%%%%%%%%%%%%%%%%%%
%%%%%%%%%%%%%%%%%%%%%%%%%%%%%%%%%%%%%%%%%

\subsection{Software Interface}
To create an interface where we can efficiently collect challenge-response pairs from the PUF, we create a block design including a Microblaze processor with our packaged PUF IP connected to it on its address bus.
\autoref{fig:vitis_block_diagram} shows our microblaze-based design.
We create an Advanced Peripheral Bus (APB) interface for our PUF to provide challenges, start trigger, and read the response.
We also add an XADC wizard to the block design to provide die temperature measurements for temperature variation testing.
The block design and compiled FPGA bitstream is used to create Xilinx Support Archive (XSA) file using which we create a Vitis platform.
We then setup a C program setup to run on the microblaze to query the PUF with challenges and collect responses, and also for tuning the tuning blocks.
After tuning, we save the log being output to the serial monitor to effectively collect the challenge-response pairs for analysis.

\subsection{Evalutated Metrics}
\label{subsec:evaluated_metrics}
We evaluated our pass-through variant PUF on seventeen (17) Arty-A7 100t FPGA boards and our permuted-variant on thirty (30) different line-seeds combinations on different boards.
We evaluated on metrics of uniformity, uniqueness, bit-aliasing, and reliability as explain in~\autoref{subsection:background}.
~\autoref{tab:pass_puf_results} shows metrics for the pass-through implementation.
\begin{table}[b]
\centering
\footnotesize
\caption{Measured PUF characteristics in Pass-Through Mode}
\label{tab:pass_puf_results}
\begin{tabular}{lcc}
\hline
\textbf{Metric} & \textbf{Measured(mean)} & \textbf{Ideal} \\
\hline
Uniformity & $49.54\%$ & $50\%$ \\
Neighbour Difference Rate & $49.11\%$ & $50\%$ \\
Uniqueness & $48.49\%$ & $50\%$ \\
Reliability & $93.62\%$ & $100\%$ \\
Reliability (Majority Voting) & $97.31\%$ & $100\%$ \\
\hline
\end{tabular}
\end{table}
Notice that we also have the reliability row with majority voting.
This is because baseline reliability was not good enough at $93.62\%$ mainly due to the XOR of four arbiters which means the errors add up.
The majority voting we implement is simply a 7-time repititive evaluation of the PUF for the same challenge, and the response is simply whether the majority evaluations were 0 or 1.
The uniformity and neighbour difference rate is tuned at time of tuning.
For the pass-through variant we were unable to get a mean uniqueness better than $\approx 40\%$ when using only a single arbiter such as $\rho_3$. 
This was most likely due to the routing asymmetry on the FPGA routing matrix which dominates the delay difference between the delay lines rather than the random routing delay variation between the boards.
For this reason, we make use of all four signals in the last delay blocks to evaluate four arbiters $\rho_{0}$, $\rho_{1}$, $\rho_{2}$ and $\rho_{3}$, and take their XOR which gives us almost ideal uniqueness.
This way, our design avoids a poor uniqueness trap by not doing any delay line crossing as in the line-crossing implementation in~\autoref{subsection:background} which adds dominant asymmetric routing which dominates the random variation across boards.
At the same time, our design includes the wiring delays in the switching rather than making it common to all paths in the delay line as in the non-crossing implementation.
This does add some asymmetric routing which reduces the effect of random variation, but we counteract its effect on uniqueness by using tuning blocks and a four arbiter XOR.
\begin{table}[t]
\centering
\footnotesize
\caption{Measured PUF characteristics in Permuted Mode}
\label{tab:permuted_puf_results}
\begin{tabular}{lcc}
\hline
\textbf{Metric} & \textbf{Measured(mean)} & \textbf{Ideal} \\
\hline
Uniformity & $48.22\%$ & $50\%$ \\
Neighbour Difference Rate & $49.48\%$ & $50\%$ \\
Uniqueness (Across different Seed-Pairs (m is variable), $\nu_{m,0} \neq \nu_{m,1}$) & $47.23\%$ & $50\%$ \\
Uniqueness (Across Boards w/ same Seed-Pairs (m is constant), $\nu_{m,0} \neq \nu_{m,1}$) & $2.11\%$ & $50\%$ \\
Reliability & $97.78\%$ & $100\%$ \\
\hline
\end{tabular}
\end{table}

Next, we evaluate our different line-seeds permuted-variant.
For each evaluation, a seed-pair $(\nu_{m,0}, \nu_{m,1})$ is assigned to the PUF or every delay line is assigned a seed $\nu_{m,j}$ where $j =\{0,1\}$ to denote the delay line and $m$ is the number of different seed pairs evaluated.
We clarify that in this paper $\nu_j$ refers to the seed for a delay line $j = \{0,1\}$ and $\nu_{m,j}$ refers to the seed assigned for the unique seed-pair instance evaluation $m$ and delay line $j$.
The important thing to understand is that the seeds for each line are different in a single evaluation $\nu_{m,0} \neq \nu_{m,1}$ to implement a different permutation, and the seed pairs across evaluations are also different so that we try a unique pair for each evaluation $\nu_{m=x, j} \neq \nu_{m\neq x, j}$.
We will clarify again that this does not constitute the traditional definition of a PUF.
As PUF instances are unique across devices, and our permuted instances are unique across $\vec c \to \vec S$ permutation in each block in the delay line.
As explained, two seeds are chosen in each instance: one for each delay line.
The seeds $\nu_{m,j}$, where $\nu_{m,0} \neq \nu_{m,1}$, then deterministically choose permutations $\pi_{j,g}$ for the $\vec c \to \vec S$ mappings of block $g$ in line $j$.
For the same seed-pairs, meaning $\nu_{m,0} \neq \nu_{m,1}$ and $m$ is constant, the PUF showed near 0 uniqueness across boards.
For different seed pairs meaning that $\nu_{m,0} \neq \nu_{m,1}$ and $m$ is variable: $[(\nu_{0,0}, \nu_{0,1}), (\nu_{1,0}, \nu_{1,1}), (\nu_{2,0}, \nu_{2,1})...]$, the PUF showed near ideal uniqueness across seed-pairs or $m$.
We can explain this by considering that for the same challenges each line chooses a different delay line path on account of different seeds and consequently different block permutations.
Therefore the difference caused by these different paths is significantly greater than any random routing delay variations across the boards:
\begin{equation}
    \Delta_{j,\nu_{m,j}} = \delta_{route}(\nu_{m,j}) + \delta_{random}(\nu_{m,j})
\end{equation}
Here $\Delta_{j, \nu_{m,j}}$ is the delay accumulated for delay line $j$ and its seed $\nu_{m,j}$.
$\delta_{route}(\nu_{mj})$ is the route taken for any particular challenge for the specific line seed $\nu_{m,j}$.
$\delta_{random}(\nu_{m,j})$ is the random puf delay variation of that route.
Subtracting the delay of the two lines:
\begin{equation}
    \Delta_{0,\nu_{m,0}} - \Delta_{1,\nu_{m,1}} = \delta_{route}(\nu_{m,0}) - \delta_{route}(\nu_{m,1}) + \delta_{random}(\nu_{m,0}) - \delta_{random}(\nu_{m,1}) 
\end{equation}
We know from prior PUF literature that routing differences are significantly greater than the random process variation, $|\delta_{route}(\nu_{m,0}) - \delta_{route}(\nu_{m,1})| >>> |\delta_{random}(\nu_{m,0}) - \delta_{random}(\nu_{m,1})|$.
Therefore, the delay difference from route chosen due to the difference in permutation mapping in blocks on account of different seeds is what uniquely decides the response for a particular challenge rather than the device the PUF is evaluated on.
In other words, the unique seed pair instance $m$ decides the difference in routing.
For constant $m$, the routing difference is the same and the random variation across boards is not sufficient to dominate the PUF response.
For variable $m$, the routing difference changes with $m$ and therefore dominates the PUF response rather than the random variation.
~\autoref{tab:permuted_puf_results} shows the results for our permuted-mode instances.
We use a single arbiter here which gives us a near ideal reliability of $97.78\%$ therefore we do not need to perform majority voiting as in our four arbiter XOR scheme for the pass-through mode.
The more interesting results are of uniqueness of across seed-pairs (variable m) which is close to ideal at $47.23\%$ and uniqueness across boards with the same seed-pair (constant m) which is nearly 0 at $2.11\%$.
These results means this mode of ours is not exactly a PUF, however, we still find it interesting to present here.
Characterizing this instance and other analysis is left as future work.

It should also be clarified that this uniqueness behavior is mainly due to the use of different seeds $\nu_0 \neq \nu_1$ in each delay line, which causes different permutation and different routing per challenge.
If the same seed for both delay lines was used, $\vec c \to \vec S$ mapping would still be permuted but the mapping would be identical for both delay lines.
This means that $\delta_{route}(\nu_0) = \delta_{route}(\nu_1)$ since $\nu_0 = \nu_1$, so the delay difference would be determined by random delay variations only assuming ideal matched routing.
In this case, the permuted mode with same seeds would be a PUF instance with optional obfuscated $\vec c \to \vec S$ mapping for each delay line challenge blocks if the seed is kept secret.
We do not evaluate this case, as we feel the pass-through mode is a similar instance with reordering of challenges, but this mode does provide the additional feature of obfuscating the challenge application to each delay block.

Further, it should be stated that all the above metrics are averages across the board and seed populations and the above metrics were obtained after tuning each board for the same FPGA bitstream.
Almost all boards required a different tuning set $T_0, T_1$ for near ideal uniformity and neighbour difference rate, however, for a single arbiter implementation with pass-through mode, uniqueness was still not above $\approx40\%$ with different tunings, which is an interesting results that shows that routing asymmetry still dominates even with different tunings sets $T_0. T_1$  for each board.
But with our four arbiter XOR, our uniqueness was near ideal for the boards.
Our tuning algorithm ensures that apart from the XOR response, all four arbiters are sufficiently uniform, have a near-50\% neighbour-difference rate and are sufficiently unique between each other as well as~\autoref{alg:puf_tuning} shows.
Even when implementing the permuted mode with the same seed-pair across different boards, tuning software produced a different tuning set for each board, but the uniqueness was still $2.11\%$ which is an interesting result that showed that the same seed-pair uniqueness across different boards can not be eliminated even with different tuning sets.

%%%%%%%%%%%%%%%%%%%%%%%%%%%%%%%%%%%%%%%%%
%%%%%%%%%%% FIGURE 6 %%%%%%%%%%%%%%%%%%%%
%%%%%%%%%%%%%%%%%%%%%%%%%%%%%%%%%%%%%%%%%
\begin{figure*}[b]
\centering
\includegraphics[width=1.0\textwidth]{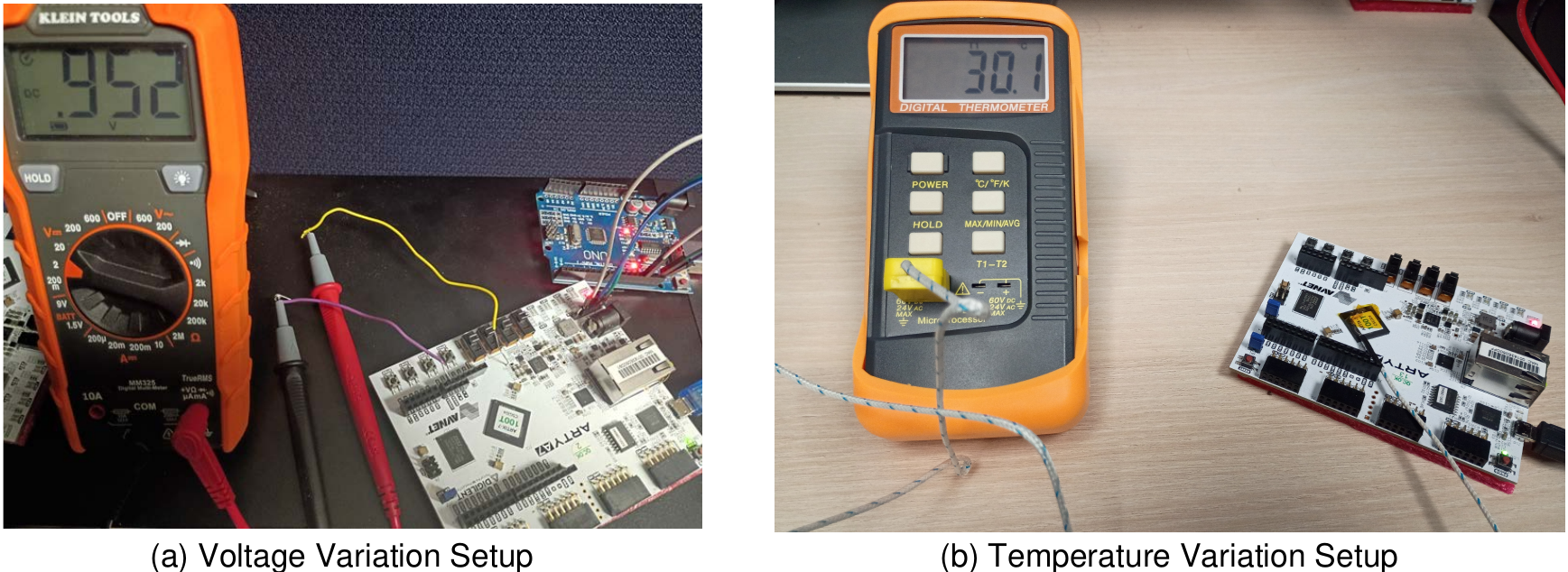}
\caption{Voltage and Temperature variation setup for Arty FPGA.}
\Description{Voltage and Temperature variation hardware setup of Arty FPGA is shown. For voltage variation, a voltmeter is show connected to FPGA pins and an Arduino connected to a header of the FPGA via jumper wires. For temperature variation, a k-type thermocouple is shown with its probe placed on the FPGA die and attached with conducting tape.}
\label{fig:volt_var}
\end{figure*}
%%%%%%%%%%%%%%%%%%%%%%%%%%%%%%%%%%%%%%%%%
%%%%%%%%%%%%%%%%%%%%%%%%%%%%%%%%%%%%%%%%%
%%%%%%%%%%%%%%%%%%%%%%%%%%%%%%%%%%%%%%%%%

\subsection{Reliability under Voltage and Temperature Variation}
Next, we will test one of our PUF instances for reliability under voltage variation and temperature variation.
This is so that we can evaluate that the PUF instance remains suitably reliable under environmental variation.
We only test the pass-through mode of our PUF for reliability here since it consists of an XOR of four arbiters to generate the response which is more prone to errors.
Our permuted mode uses just one of those four arbiters as the response so it can be safely assumed that reliability of the XOR can be at best just as good as the reliability of a single arbiter.

First we will go over our voltage variation setup.
We vary the FPGA core supply rail $V_{CCINT}$ by reprogramming the onboard DA9062 power-management IC.
The Arty A7-100T employs an XC7A100TCSG324-1 FPGA and the nominal $V_{CCINT}$ recommended by AMD/Xilinx for this device is $1.00V$, with a recommended operating range of $0.95V$--$1.05V$.
We make use of the onboard DA9062 PMIC, which the Arty uses to generate $V_{CCINT}$.
The $J11$ header provided next to the barrel-jack adapter exposes the PMIC $I^2C$ interface.
We use an Arduino Uno as an external $I^2C$ controller to program the PMIC through the $J11$ $I^2C$ interface and vary the voltage of the $V_{CCINT}$ rail.
The Arty remains connected to the host computer through its normal Micro-USB interface for FPGA programming, MicroBlaze execution, and PUF data acquisition.
We use a voltmeter to measure the rail voltage in real time at the $V_{CCINT}$ power path near resistor $R252$ provided on the board PCB.
\autoref{fig:volt_var} shows our setup.
The Arty is uploaded with a sketch that tests a software-tuned HAPUF instances reliability over $5,000$ different challenges with each challenge evaluated $50$ times.
We measure XOR reliability without majority voting and with majority voting with our majority voting scheme taking a majority vote over 7 repititions.
%%%%%%%%%%%%%%%%%%%%%%%%%%%%%%%%%%%%%%%%%
%%%%%%%%%%% FIGURE 7 %%%%%%%%%%%%%%%%%%%%
%%%%%%%%%%%%%%%%%%%%%%%%%%%%%%%%%%%%%%%%%
\begin{figure*}[t]
\centering
\includegraphics[width=\textwidth]{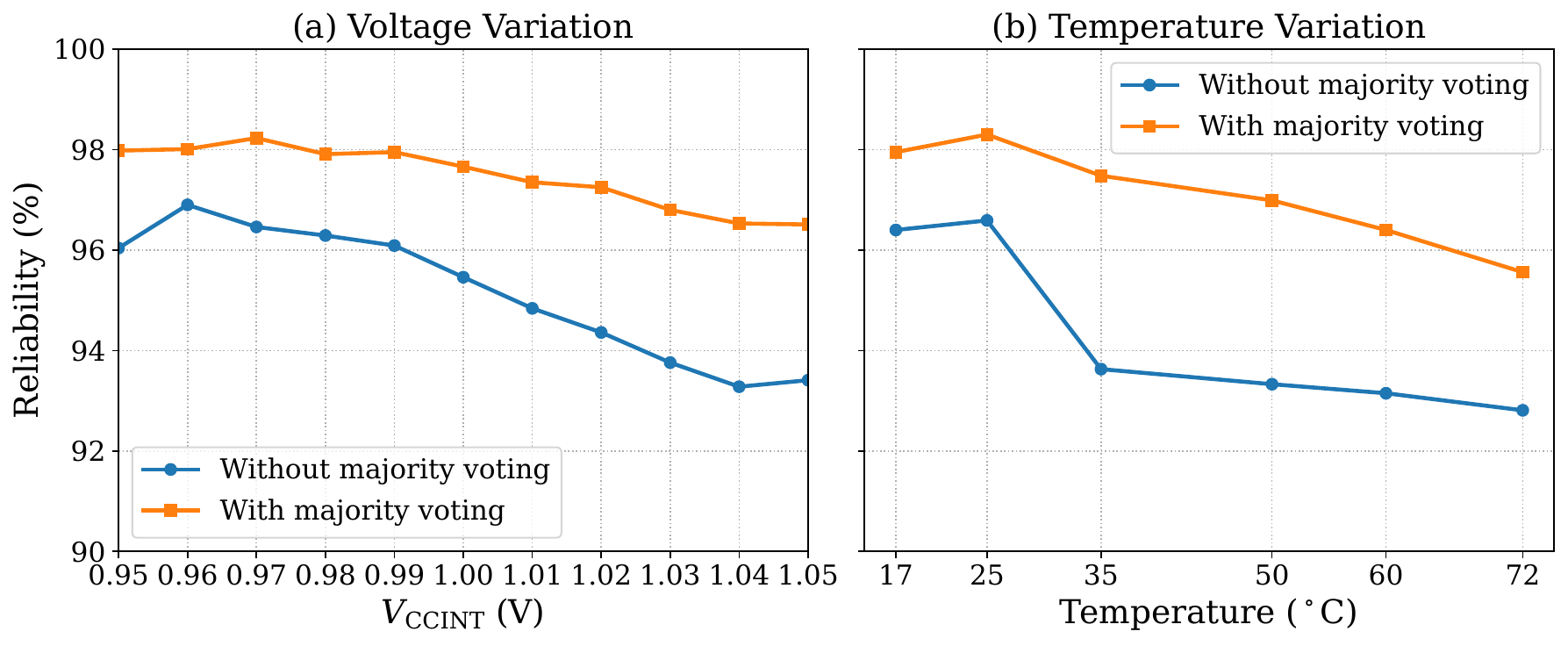}
\caption{Reliability against Voltage and Temperature Variation.}
\Description{Two graphs show the variation of reliability with voltage and temperature respectively. Both graphs show good reliability and the lower extremes and show decreasing reliability as voltage or temperature increases.}
\label{fig:reliability_variation}
\end{figure*}
%%%%%%%%%%%%%%%%%%%%%%%%%%%%%%%%%%%%%%%%%
%%%%%%%%%%%%%%%%%%%%%%%%%%%%%%%%%%%%%%%%%
%%%%%%%%%%%%%%%%%%%%%%%%%%%%%%%%%%%%%%%%%

Next we explain our temperature variation setup.
As~\autoref{fig:vitis_block_diagram} shows, we include the XADC wizard in our block diagram.
This provides a temperature sensor for the FPGA die that we can poll in our Vitis program.
The FPGA is placed close to a heat or cooling source, and our Vitis program polls the XADC at intervals to see if the temperature reaches the desired level.
Once the temperature reaches the desire level for example, $50^\circ \mathrm{C}$, the program exits the while loop and then runs a reliability test.
During the reliability test, the XADC is still polled to check if the FPGA does not drop below the deired level beyond a set tolerance such as $\pm 1^\circ \mathrm{C}$.
Further, we attach a fine-wire k-type thermocouple sensor to the die to provide another check on the temperature readings as~\autoref{fig:volt_var} shows, but we believe that the XADC is the best way to know the real-time temperature while doing variation testing.
As in the voltage-variation case, we test normal reliability and reliability with majority voting.
The Arty is again uploaded with a sketch that tests a software-tuned HAPUF instances reliability over $5,000$ different challenges with each challenge evaluated $50$ times.

\autoref{fig:reliability_variation} shows the results of our experiments.
Although our experiments show that reliability changes with voltage and temperature variation, we have some untraditional results of reliability slight improving at lower temperature and voltage.
We do not have a proven explanation for this, but it can be hypothesized that for this particular PUF instance, the tuning might have balanced the delay lines so that at lower temperatures and voltage, the random variation become significant enough to dominate the result and improve reliability.

\subsection{Comparison with other Delay Blocks}
In this section we will show how our implementation compares with other delay blocks proposed in previous literature.
We will implement the standard non-crossing and crossing implementations as shown in~\autoref{subsection:background} and covered in literature by~\cite{majzoobi2010fpga} and ~\cite{lim2005_affine} respectively.
Other than this, we will implement an APUF using the Path-Changing-Switch (PCS) implemented by Mahalat et al.~\cite{mahalat2021PCS}.
Their delay block is a 4x1 switch implemented using a complete FPGA slice.
Finally, we will also implement the Flip-Flop APUF (FF-APUF) using the delay block proposed in~\cite{gu2019ffapuf}.
The FF-APUF also uses a 4x1 switch for a delay block but uses Flip-Flops instead of LUTs as delay elements.
All designs are implemented to have at least 10 or 20 delay blocks with the aim of having a total of 20 challenge bits.
The PCS-APUF and FF-APUF have two challenges per delay block so we implement 10 delay blocks per delay line.

All four designs are constrained with BEL, LOC and LOCK\_PINS constraints to make the delay lines symmetric and identical.
In modern FPGA implementation flows, it is often not possible to constrain the routing of two delay lines to be symmetric.
Even though the placement of resources in the two delay lines is identical and symmetric using BEL and LOC constraints, and LOCK\_PINs constraints are used to further ensure identical primitives block pins are also used, the routing resources in the FPGA matrix are not necessarily identical in two slice columns.
For this reason, the routing constraint applied to one delay line may be rejected by the router in another delay line.
Or if another unrelated route exists that uses one of the identical routing resource that the other delay line uses, the same routing constraint for the delay line will be rejected on account of the routing resource having already been used. 
This can require unrouting some nets and routing the PUF nets first.
In previous literature, hard macros were utilized to create a copy of an implemented delay line on another slice column with symmetric and identical placement and routing, however, modern relatively placed macros (RPMs) do not necessarily copy over identical routing to the copied delay line.
Therefore, for our HAPUF implementation, and these other four comparisons, we attempt to constrain routing or re-run the router to the point that the delay of both lines is balanced, but we do not do thorough routing that is an exact replication of how routing is proposed in the relevant literature for these designs.
While the use of BEL, LOC, and LOCK\_PINs constraints are relatively straightforward for an experienced user, routing constraints requires trial and error, repetition and knowledge of the routed design and FPGA routing matrix which is sometimes not available to the user.
Further, a certain routing constraint may produce a uniform design on one FPGA but a non-uniform one in another.
For this reason we treat all designs with the same tuning based treatment without thorough routing to provide a comparison to our ease-of-implementation claim.

The results for our HAPUF are shown in~\autoref{subsec:evaluated_metrics} for a tuned implementation with alternately placed tuning blocks.
For our four comparison implementations, we compare all four with three different configs: no tuning blocks, challenge blocks followed by tuning blocks in the end, and alternately placed tuning blocks.
\autoref{tab:puf_comparison} shows the results of our other four comparison implementations in their three variants.
\begin{table*}[t]
\footnotesize
\centering
\caption{Comparison of the four PUF variants under different tuning configurations.}
\setlength{\tabcolsep}{3.5pt}
\begin{tabular}{p{3.0cm} p{3.0cm} c c c c}
\hline
\textbf{PUF Variant} &
\textbf{Configuration} &
\textbf{Uniformity (\%)} &
\textbf{NDR (\%)} &
\textbf{Uniqueness (\%)} &
\textbf{Reliability (\%)} \\
\hline

\multirow{3}{3.0cm}{Standard APUF (Non-crossing Implementation)}
& No Tuning              & 0.00 & 0.00 & N/A & N/A \\
& Line-End Tuning Blocks & 0.00 & 0.00 & N/A & N/A \\
& Alternate Tuning Blocks& 0.00 & 0.00 & N/A & N/A \\
\hline

\multirow{3}{3.0cm}{Standard APUF (Crossing Implementation)}
& No Tuning              & 59.34 & 45.64 & 3.44 & 99.57 \\
& Line-End Tuning Blocks & 52.76 & 49.31 & 4.62 & 99.23  \\
& Alternate Tuning Blocks& 49.70 & 50.57 & 2.23 & 99.81 \\
\hline

\multirow{3}{3.0cm}{PCS-APUF (Non-Crossing)}
& No Tuning              & 7.89 & 13.78 & 2.26 & 99.92 \\
& Line-End Tuning Blocks & 51.11 & 49.26 & 3.75 & 99.60 \\
& Alternate Tuning Blocks& 52.56 & 49.41 & 8.27 & 99.00 \\
\hline

\multirow{3}{3.0cm}{FF-APUF (Non-Crossing)}
& No Tuning              & 18.10 & 34.45 & 6.08 & 99.23 \\
& Line-End Tuning Blocks & 50.85 & 50.34 & 11.10 & 97.94 \\
& Alternate Tuning Blocks& 48.81 & 49.16 & 8.83 & 99.15 \\
\hline

\multirow{1}{3.0cm}{HAPUF (Pass-Through)}
& Alternate Tuning Blocks& 49.54 & 49.11 & 48.49 & 97.31 \\
\hline

\multirow{1}{3.0cm}{HAPUF (Permuted)}
& Alternate Tuning Blocks& 48.22 & 49.48 & 47.23* & 97.78 \\
\hline

\end{tabular}
\begin{minipage}{\linewidth}
\footnotesize
$^{*}$ The HAPUF permuted variant uniqueness is across seed pairs with variable $m$ where $\nu_{m,0} \neq \nu_{m,1}$ and not across boards like the other entries in the table. 
\end{minipage}
\label{tab:puf_comparison}
\end{table*}
The table also has entries for the two HAPUF variants.
The statistics are the same as those in~\autoref{subsec:evaluated_metrics} but shown here again for better readability for comparing.
Only for the standard non-crossing implementation of the APUF were we unable to get good metrics in uniformity and neighbour difference rate.
This can be explained due to the large bias term as discussed in~\autoref{subsection:background}.
Eliminating this term requires controlled and careful routing which we forego in favour of an easier software based tuning. 
Although the other PUF implementation also did not have good metrics for the no-tuning configs, the software based-tuning shows promise here as their tuned versions have near ideal uniformity and neighbor difference rates.
Additionally, tuning configuration or the placement of tuning blocks can be decided at runtime through software as discussed before.
However, all comparison implementations perform poorly in uniqueness.
This can be due to the fact that it is difficult to eliminate the asymmetric routing in FPGAs so that the asymmetric routing delays do not dominate the results.
This will require repetitive and thorough routing control and as our experiment shows, tuning is not enough for these delay blocks to improve uniqueness as we evaluate these results on different boards with different tuning sets.
Therefore, to the best of our knowledge and from our results, our HAPUF design using our novel delay structure using both the pass-through and permuted variant provides near ideal PUF metrics with a single automated software-based tuning run as compared to thorough manual routing.
We do note that for our pass-through variant, without a four-arbiter XOR, a single arbiter implementation had uniqueness at most of 40\%, so our design also suffers from asymmetric routing.
But since our design has four outputs from the delay lines for the same resource cost of a complete slice, we can place four arbiters without any extra resource cost to improve uniqueness with a XOR variant.
The PCS-APUF and FF-APUF variants would require four times the hardware resources to implement such a variant.
Further, our different line-seeds permuted versions of HAPUF provide near ideal uniqueness across unique different seed-pairs with a single arbiter.

%%%%%%%%%%%%%%%%%%%%%%%%%%%%%%%%%%
%%%%%%%%%%%%%%%%%%%%%%%%%%%%%%%%%%
%%%%%%%%%%%%%%%%%%%%%%%%%%%%%%%%%%

%%%%%%%%%%%%%%%%%%%%%%%%%%%%%%%%%%
%%%%%%%%%%% SECTION V %%%%%%%%%%%%
%%%%%%%%%%%%%%%%%%%%%%%%%%%%%%%%%%
\section{Conclusion}
\label{sec:Conclusion}
In this paper, we address the need for a new APUF design that provides an automated way to implement an APUF variant on FPGA boards without repetitive trial-and-error manual routing with acceptable PUF metrics.
We achieve this with multiple contributions.
First, we propose a novel delay block design using LUTs and carry-chains within a slice which can be operated in pass-through and permuted modes.
Next, we arrange these four-input four-output delay blocks to propose our novel HAPUF design.
Finally, we make use of software automated tuning to find a tuning configuration that provides near-ideal PUF metrics with a single software tuning run.
Although manual routing is an established method of implementing PUFs on FPGAs, we provide a software-based alternative for implementation with the motivation of an easier software tuning run rather than design implementation and routing repetitions.
Some limitations exist.
All of our experiments involve pre-processing of the challenge to distribute the challenge bits.
This improves the neighbor difference rate in our and in the comparison implementations, but poor neighbor difference rate with sequential challenges is a known issue in arbiter PUFs due to their linear delay models.
This is not a limitation unique to our design; it is a known feature of all APUF-based designs.
Finally, we acknowledge that with manual routing, our comparison implementations have ideal PUF metrics as reported in PUF literature, however, we show that our novel implementation, motivated by design choices from previous literature, provides an alternate software automated method to tune PUF instances to provide near-ideal metrics without the strict manual routing required in literature.   

% \section*{Acknowledgment}
% This material is based upon work supported by the U.S. Department of Energy, Office of Cybersecurity, Energy Security, and Emergency Response (CESER), under Award Number DE-CR0000050. This report was prepared as an account of work sponsored by an agency of the United States Government. Neither the United States government nor any agency thereof, nor any of their employees, makes any warranty, express or implied, or assumes any legal liability or responsibility for the accuracy, completeness, or usefulness of any information, apparatus, product, or process disclosed or represents that its use would not infringe privately owned rights. Reference herein to any specific commercial product, process, or service by trade name, trademark, manufacturer, or otherwise does not necessarily constitute or imply its endorsement, recommendation, or favoring by the United States Government or any agency thereof. The views and opinions of authors expressed herein do not necessarily state or reflect those of the United States Government or any agency thereof.

% \end{thebibliography}
\bibliographystyle{IEEEtran}
\bibliography{references}

\end{document}